\documentclass[12pt]{article}
\usepackage{epsf,a4,latexsym}
\usepackage[small,bf]{caption}

\newcommand{\sixth}{\mbox{\small $\frac{1}{6}$}}         
\newcommand{\half}{\mbox{\small $\frac{1}{2}$}}          
\newcommand{\third}{\mbox{\small $\frac{1}{3}$}}         
\newcommand{\fivesixth}{\mbox{\small $\frac{5}{6}$}}     
\newcommand{\ti}{\mbox{\tiny $T\!I$}}                    
\newcommand{\lat}{\mbox{\tiny $L\!A\!T$}}                
\newcommand{\WI}{\mbox{\tiny $WI$}}                      
\newcommand{\plaquette}{\mbox{\tiny $Plaquette$}}        
\newcommand{\rectangle}{\mbox{\tiny $Rectangle$}}        
\newcommand{\QCD}{\mbox{\tiny $Q\!C\!D$}}                

\def\lsim{\mathrel{\rlap{\lower4pt\hbox{\hskip1pt$\sim$}}
    \raise1pt\hbox{$<$}}}                
\def\gsim{\mathrel{\rlap{\lower4pt\hbox{\hskip1pt$\sim$}}
    \raise1pt\hbox{$>$}}}                
\begin{document}

\title{
\vspace{-3.0cm}
\flushright{\normalsize DESY 08-141} \\
\vspace{-0.35cm}
{\normalsize Edinburgh 2009/01} \\
\vspace{-0.35cm}
{\normalsize LTH 819} \\
\vspace{-0.35cm}
{\normalsize January 2009} \\
\vspace{0.5cm}
\centering{\Large \bf Non-perturbative improvement of stout-smeared
                      three flavour clover fermions}}

\author{\large
        N.~Cundy$^{a}$, M.~G\"ockeler$^{a}$,
        R.~Horsley$^{b}$, T.~Kaltenbrunner$^{a}$, \\
        A.~D. Kennedy$^{b}$, Y.~Nakamura$^{ac}$,
        H.~Perlt$^{d}$, D.~Pleiter$^c$, \\
        P.~E.~L.~Rakow$^e$, A.~Sch\"afer$^{a}$,
        G.~Schierholz$^{af}$, A.~Schiller$^{d}$, \\
        H.~St\"uben$^g$ and J.~M.~Zanotti$^b$ \\[1em]
         -- QCDSF-UKQCD Collaboration -- \\[1em]
        \small $^a$ Institut f\"ur Theoretische Physik,
               Universit\"at Regensburg, \\[-0.5em]
        \small 93040 Regensburg, Germany \\[0.25em]
        \small $^b$ School of Physics and Astronomy,
               University of Edinburgh, \\[-0.5em]
        \small Edinburgh EH9 3JZ, UK \\[0.25em]
        \small $^c$ Deutsches Elektronen-Synchrotron DESY, \\[-0.5em]
        \small 15738 Zeuthen, Germany \\[0.25em]
        \small $^d$ Institut f\"ur Theoretische Physik,
               Universit\"at Leipzig, \\[-0.5em]
        \small 04109 Leipzig, Germany \\[0.25em]
        \small $^e$ Theoretical Physics Division,
               Department of Mathematical Sciences, \\[-0.5em]
        \small University of Liverpool,
               Liverpool L69 3BX, UK \\[0.25em]
        \small $^f$ Deutsches Elektronen-Synchrotron DESY, \\[-0.5em]
        \small 22603 Hamburg, Germany \\[0.25em]
        \small $^g$ Konrad-Zuse-Zentrum
               f\"ur Informationstechnik Berlin, \\[-0.5em]
        \small 14195 Berlin, Germany }

\date{}

\maketitle


\begin{abstract}
We discuss a $3$-flavour lattice QCD action with clover improvement in
which the fermion matrix has single level stout smearing for the hopping
terms together with unsmeared links for the clover term. With the (tree-level)
Symanzik improved gluon action this constitutes the {\bf S}tout {\bf Li}nk
{\bf N}on-perturbative {\bf C}lover or SLiNC action.
To cancel $O(a)$ terms the clover term coefficient has to be tuned.
We present here results of a non-perturbative determination of this
coefficient using the Schr\"odinger functional and
as a by-product a determination of the critical hopping parameter.
Comparisons of the results are made with lowest order perturbation theory.

\end{abstract}

\clearpage


\section{Introduction and $O(a)$ improvement}
\label{improvement}


When constructing a lattice QCD action, even the simplest gluon
action has only $O(a^2)$ corrections. The naive quark action also has
$O(a^2)$ corrections, but suffers from the `doubling problem' describing
$16$ flavours in the continuum limit. A cure is to add the Wilson mass term,
so $15$ flavours decouple in the continuum limit, but the price is that
there are now $O(a)$ corrections (and also loss of chiral invariance),
so that for example for a ratio of hadron masses
\begin{eqnarray}
   {m_H \over m_{H^\prime}} = r_0 + ar_1 + O(a^2) \,.
\end{eqnarray}
The Symanzik approach is a systematic improvement to $O(a^n)$ 
(where in practice $n=2$ for the fermion action) by adding a basis
(an asymptotic series) of irrelevant operators and tuning their
coefficients to remove completely $O(a^{n-1})$ effects.
Restricting improvement to on-shell quantities the equations
of motion reduce the set of operators in both the action and
in matrix elements. Indeed, for $O(a)$ improvement of the fermion
action only one additional flavour-singlet operator is required
\begin{eqnarray}
  {\cal L}_{clover} \propto a c_{sw}
     \sum_{q,x,\mu\nu} \overline{q}(x) \sigma_{\mu\nu}F_{\mu\nu}(x)q(x) \,,
\end{eqnarray}
the so-called  `Sheikholeslami--Wohlert' or `clover' term,
\cite{sheikholeslami85a}. So if we can improve {\it one} on-shell
quantity this then fixes $c_{sw}$ as a function of the lattice spacing
$a$ or equivalently of the bare coupling $g_0^2$, so that all other
on-shell quantities are automatically improved to $O(a)$,
i.e., we now have
\begin{eqnarray}
  {m_H \over m_{H^\prime}} = r_0 + O(a^2) \,.
\end{eqnarray}
A non-perturbative determination of $c_{sw}$ will be the main goal of
this paper, the general approach being described below.

Matrix elements still require additional $O(a)$ operators, for example
for the axial current and pseudoscalar density, \cite{luscher96a}%
\footnote{We implicitly distinguish between quark flavours in
operators, i.e.\ consider non-singlet operators.},
\begin{eqnarray}
   {\cal A}_\mu &=& ( 1 + b_Aam_q )( A_\mu + c_A a\partial^{\lat}_\mu P )
                                           \nonumber \\
   {\cal P}     &=& ( 1 + b_Pam_q ) P \,,
\end{eqnarray}
(for mass degenerate quarks) with
\begin{eqnarray}
   A_\mu = \overline{q}\gamma_\mu\gamma_5 q \,, \qquad 
   P     = \overline{q}\gamma_5 q \,,
\end{eqnarray}
which require additional $b_A$, $c_A$ and $b_P$ improvement coefficients.
An easily determined quantity is the quark mass computed from
the PCAC WI relation%
\footnote{$\partial^{\lat}_\mu$ is the symmetric lattice derivative,
$(\partial^{\lat}_\mu f)(x) = [ f(x+a\hat\mu) - f(x-a\hat\mu) ] /(2a)$,
and (no $\mu$ summation), $(\partial_\mu^{2\,\lat} f)(x) =
[ f(x+a\hat\mu) - 2f(x) + f(x-a\hat\mu) ] / a^2
= (\partial_\mu^{\lat}\partial_\mu^{\lat} f)(x) + O(a^2)$.},
\begin{eqnarray}
   m_q^{\WI} = { \langle 
                \partial^{\lat}_0 ( A_0(x_0) + c_A a\partial^{\lat}_0 P(x_0)) O
                 \rangle
                 \over 2 \langle P(x_0) O \rangle } \,.
\label{mq_WI_def}
\end{eqnarray}
Choosing different operators, $O$, gives different determinations
of the quark mass $m_{q}^{\WI\,(i)}$, $i = 1$, $2$ with different
lattice artifacts. If the quark mass is improved then its errors
are $O(a^2)$. So we can determine the `optimal' $c_{sw}$ improvement
coefficient by tuning until
\begin{eqnarray}
   m_{q}^{\WI\,(1)} = m_{q}^{ \WI\,(2)} \,.
\label{mqWI_equality}
\end{eqnarray}
(This is equivalent to considering the renormalised quark mass 
\begin{equation}
   m_{qR} = {Z_A (1 + b_Aam_q) \over Z_P(1 + b_Pam_q)} \, m_q^{\WI} \,.
\end{equation}
In general the $b_A$, $b_P$ coefficients do affect considerations
of $O(a)$-improvement. However, here one imposes a condition at fixed
bare parameters $(g_0^2, m_q)$ which means that the factors drop out.)
Practically, how this is achieved will be discussed in this paper
after the action is introduced.

This paper is organised as follows. In the next section, section~\ref{slinc},
the action is given and in the following section the Schr\"odinger
functional is briefly discussed, together with the general procedure for
determining the optimal $c_{sw}$ and optimal critical hopping parameter,
$\kappa_c$. Section~\ref{lattice} gives some lattice details for a
series of simulations at various coupling constants, which after suitable
interpolations leads to this determination.
Section~\ref{finite_size_effects} then discusses possible finite
size effects in the results. Results are collected together in
section~\ref{results} and a polynomial interpolation (in the coupling
constant) for both $c_{sw}$ and $\kappa_c$ are given, together with
a comparison with the lowest order perturbation result. Finally in 
section~\ref{conclusions} some brief conclusions are discussed. Tables
of the raw results are given in appendix~\ref{appendix_raw_results}.


\section{The SLiNC action}
\label{slinc}


We shall consider here $n_f = 3$ flavour stout link clover fermions
-- SLiNC fermions (Stout Link Non-perturbative Clover). In a little more
detail, for each flavour,
\begin{eqnarray}
   \lefteqn{S_F =}
                                                \nonumber              \\
   && \sum_x \left\{
      \kappa \, \sum_{\mu}[\overline{q}(x)(\gamma_\mu - 1)
                         \tilde{U}_\mu(x) q(x+a\hat{\mu}) -
             \, \overline{q}(x)(\gamma_\mu + 1) 
                         \tilde{U}^\dagger_\mu(x-a\hat{\mu}) q(x-a\hat{\mu})]
                  \right.
                                                \nonumber              \\
   && \hspace*{0.75in} \left.
            + \overline{q}(x)q(x) -
             \half \kappa a c_{sw} \sum_{\mu\nu}
                 \overline{q}(x)\sigma_{\mu\nu}F_{\mu\nu}(x)q(x)
                            \right\} \,.
\label{action}
\end{eqnarray}
Rescaling the quark fields $q \to q/\sqrt{2\kappa}$ gives the
quark mass $m_q$ where
\begin{equation}
   m_q(c_{sw}) = {1 \over 2a} \left(
                             {1 \over\kappa} - {1 \over \kappa_c(c_{sw})}
                      \right) \,,
\label{VWI_qm}
\end{equation}
which is proportional to the PCAC quark mass, $m_q^{\WI}$. The loss of
chiral invariance means that for a given $c_{sw}$ a critical hopping
parameter, $\kappa_c(c_{sw})$ has now also to be determined.

The hopping terms (Dirac kinetic term and Wilson mass term,
i.e.\ those terms involving a $\kappa$) in eq.~(\ref{action})
use a once stout smeared link or `fat link', \cite{morningstar03a},
\begin{eqnarray}
   \tilde{U}_\mu
            &=& \exp\{iQ_\mu(x)\}\, U_\mu(x)
                                                            \nonumber \\
   Q_\mu(x) &=& {\alpha \over 2i} \left[ VU^\dagger - UV^\dagger 
                         - \third \mbox{Tr} (VU^\dagger - UV^\dagger) \right] \,,
\end{eqnarray}
($V_\mu$ is the sum of all staples around $U_\mu$)
while the clover term remains built from `thin' links -- they are
already of length $4a$ and we want to avoid the fermion matrix 
becoming too extended. Smearing is thought to help at present lattice
spacings by smoothing out fluctuations in the gauge fields slightly
and so reducing the condition number and also to avoid a near
first order phase transition.
The critical kappa in eq.~(\ref{VWI_qm}) corresponds to
an additive mass renormalisation
\begin{equation}
   m_c(c_{sw}) = {1 \over 2a} \left(
                         {1 \over \kappa_c(c_{sw})} - {1 \over 1/8}
                      \right) \,.
\end{equation}
It is known that with a combination of link fattening and
increase of the clover coefficient, it is possible to reduce
this mass term \cite{capitani06a,degrand98a,boinepalli04a}.
The stout variation is also analytic which means that the derivative
in the gauge group can be taken (so the force in the Hybrid Monte Carlo,
or HMC, simulation is well defined) and perturbative expansions are
also possible, \cite{perlt08a}.

To complete the action we also use the Symanzik tree--level gluon action
\begin{eqnarray}
   \lefteqn{S_G =}
                                                                       \\
      &&  {6 \over g_0^2} \, \left\{
            c_0 \sum_{\plaquette} {1 \over 3} \mbox{Re\,Tr}
                ( 1 - U_{\plaquette}) +
            c_1 \sum_{\rectangle} {1 \over 3} \mbox{Re\,Tr}
                ( 1 - U_{\rectangle})
                            \right\} \,,
                                                            \nonumber
\label{symanzik_gluon}
\end{eqnarray}
together with
\begin{eqnarray}
   c_0 = {20 \over 12} \,, \,\,\, c_1 = - {1 \over 12}
              \qquad \mbox{and} \qquad
   \beta = {6c_0\over g_0^2} = {10 \over g_0^2} \,.
\end{eqnarray}
While this gluon action has elements of higher order improvement,
namely $O(a^4)$, this is not the reason that it is used here.
(The best we can hope for the fermion action is $O(a^2)$ improvement.)
Again we wish to move the action away from a nearby first-order phase
transition occuring when using the standard Wilson action
(i.e.\ $c_0 \to 1$, $c_1 \to 0$), \cite{aoki04a} by using a slightly
extended action. Different values of $c_0$ and $c_1$ can be and
have been used in the literature to address this problem, e.g.\ \cite{aoki04a}.


\section{The Schr\"odinger functional}
\label{schrodinger}


The ALPHA Collaboration determined the improvement coefficients by means
of the `Schr\"odinger functional',
\cite{luscher92a,sint93a,sint95a,luscher96a}.
Some numerical results for $c_{sw}$ for the quenched case ($n_f =0$) were
given in \cite{luscher96b,edwards97a}, for $n_f = 2$ flavours in
\cite{jansen98a} and for $n_f = 3$ flavours in
\cite{yamada04a,aoki05a,edwards08a}.
In this approach Dirichlet boundary conditions are applied on the
time boundaries to the fields. For the gluon fields, fixing them
on the boundary is then equivalent to inducing some classical
background field about which they fluctuate. It is simplest to
consider spatially constant colour diagonal fields,
corresponding to a constant chromo-electric background field.
Concretely, we consider a $L^3\times T$
lattice (with $T = 2L$) and take the background field to be
\begin{eqnarray}
   U^{c}_0(\vec{x}, x_0) &=& 1
                                           \nonumber \\
   U^{c}_k(\vec{x}, x_0) &=&  
       \exp \left( -i{a \over T}[ x_0C^{(2)} + (T-x_0)C^{(1)}] \right) \,,
\label{Uclassical}
\end{eqnarray}
with
\begin{eqnarray}
   C^{(i)} = { 1 \over L}
             \left( \begin{array}{ccc}
                      \phi^{(i)}_1 & 0           & 0      \\
                       0          & \phi^{(i)}_2 & 0      \\
                       0          & 0           & \phi^{(i)}_3 \\
                    \end{array} \right) \,,
\end{eqnarray}
and
\begin{eqnarray}
   (\phi^{(1)}_1, \phi^{(1)}_2, \phi^{(1)}_3) = (-\sixth\pi, 0, \sixth\pi) \,,
   \qquad
   (\phi^{(2)}_1, \phi^{(2)}_2, \phi^{(2)}_3) 
           = (-\fivesixth\pi, \third\pi, \half\pi) \,,
\end{eqnarray}
and fix the boundary values a posteriori. As we have an extended gauge
action (rather than the simpler Wilson gluon action), we 
fix two values at each double boundary layer and so we choose,
following \cite{klassen97a}%
\footnote{An alternative procedure using single layer boundaries is
given in \cite{aoki98a}.},
$U^c_\mu$ from eq.~(\ref{Uclassical})
at $x_0 = -a$, $0$ (lower boundary) and similarly $U^c_\mu$ at $x_0 = T-a$
and $T$ (upper boundary). The `bulk' of the lattice is thus from $x_0 = 0$
to $x_0 = T-a$. Additionally the weight factors for the gluon loops in 
eq.~(\ref{symanzik_gluon}) must be appropriately chosen on the boundary
for $O(a)$-improvement. Classically these weight factors are not difficult
to find, however a full non-perturbative determination would be difficult.
But away from the boundaries, they only affect the local
PCAC relation to $O(a^2)$ and so are not essential for the determination
of the optimal $c_{sw}$, and so it is sufficient to use the classical
values.

The fixed boundary quark fields, $ \rho, \overline{\rho}$
(taken as zero here) make simulations with $m_q \sim 0$ with no
zero mode problems possible. They are specified on the lower inner
boundary and upper inner boundary from
\begin{eqnarray}
   P^+_0 q(\vec{x}, 0) &=& \rho^{(1)}(\vec{x})
                                                             \nonumber \\
   \overline{q}(\vec{x}, 0)P^-_0
                            &=& \overline{\rho}^{(1)}(\vec{x})
                                                             \nonumber \\
   P^-_0 q(\vec{x}, T-a) &=& \rho^{(2)}(\vec{x})
                                                             \nonumber \\
   \overline{q}(\vec{x}, T-a)P^+_0
                            &=& \overline{\rho}^{(2)}(\vec{x}) \,,
\end{eqnarray}
where $P_0^\pm$ is the projection operator defined by
\begin{eqnarray}
   P_0^\pm = \half \left( 1 \pm \gamma_0 \right) \,.
\end{eqnarray}
These projections are necessary for consistency.
$ \rho, \overline{\rho}$ can be taken as sinks and sources respectively
to build operators for correlation functions. For example here we can take
at the lower inner boundary $x_0 = 0$ ($i=1$) and upper inner boundary
$x_0 = T-a$ ($i=2$) the operators
\begin{eqnarray}
   O^{(i)} = \sum_{\vec{y},\vec{z}} \,
               \left( - {\delta \over \delta \rho^{(i)}(\vec{y})} \right)
               \gamma_5
               \left( {\delta \over \delta \overline{\rho}^{(i)}(\vec{z})}
               \right) \,.
\end{eqnarray}
So we can investigate PCAC behaviour at different distances from the
boundaries.

In a little more detail, following eq.~(\ref{mq_WI_def}), we first set
\begin{eqnarray}
   r^{(i)}(x_0) = { \partial^{\lat}_0 f^{(i)}_A(x_0)
                                          \over 2 f^{(i)}_P(x_0) } \,, \qquad
   s^{(i)}(x_0) = a{ \partial^{2\lat}_0 f^{(i)}_P(x_0) \over 2 f^{(i)}_P(x_0) }
                                           \,,
\end{eqnarray}
where
\begin{eqnarray}
   f^{(1)}_A(x_0)   &=& - {1 \over n_f^2-1} \langle A_0(x_0) O^{(1)} \rangle 
                                                              \nonumber  \\
   f^{(2)}_A(T-x_0) &=& + {1 \over n_f^2-1}\langle A_0(x_0) O^{(2)} \rangle \,,
\end{eqnarray}
and
\begin{eqnarray}
   f^{(1)}_P(x_0)   &=& - {1 \over n_f^2-1} \langle P(x_0) O^{(1)} \rangle
                                                              \nonumber  \\
   f^{(2)}_P(T-x_0) &=& - {1 \over n_f^2-1} \langle P(x_0) O^{(2)} \rangle  \,.
\end{eqnarray}
Then redefine the quark mass slightly, but which coincides to $O(a^2)$
for the improved theory
\begin{eqnarray}
   M^{(i)}(x_0, y_0) = r^{(i)}(x_0) + \widehat{c}_A(y_0) s^{(i)}(x_0) \,, \qquad
   \widehat{c}_A(y_0) = - { r^{(1)}(y_0) - r^{(2)}(y_0) \over
                                     s^{(1)}(y_0) - s^{(2)}(y_0) } \,,
\end{eqnarray}
which eliminates the unknown $c_A$ in the determination of the
quark mass, \cite{luscher96b} and replaces it by an estimator,
$\widehat{c}_A$. Improvement is defined when
\begin{eqnarray}
   (M, \Delta M) = (0, 0) \,,
\label{improvement_condition}
\end{eqnarray}
where
\begin{eqnarray}
   M \equiv M^{(1)} \,, \qquad  \Delta M \equiv M^{(1)} - M^{(2)} \,,
\end{eqnarray}
are chosen at some suitable $x_0$, \cite{luscher96b}. This gives the
required optimal $c_{sw}$ and $\kappa_c$, which we will denote by
a star: $c_{sw}^*$ and $\kappa_c^*$. Conventionally, we choose
\begin{eqnarray}
   M \equiv M^{(1)}(T/2, T/4) \,, \qquad  
   \Delta M \equiv M^{(1)}(3T/4, T/4) - M^{(2)}(3T/4, T/4) \,.
\label{M+dM_def}
\end{eqnarray}

There are small changes due to the finite volume used,
so eq.~(\ref{improvement_condition}) becomes
\begin{eqnarray}
   (M, \Delta M) = (0, \Delta M^{tree}) \,,
\end{eqnarray}
where $\Delta M^{tree}$ is the tree-level (i.e.\ $g_0^2 =0$,
$c_{sw} \equiv c_{sw}^{tree} = 1$) value of $\Delta M|_{M=0}$
on the $L^3\times T$ lattice. This ensures that $c_{sw} \to 1$
exactly as $\beta \to \infty$. For $\alpha = 0$, the analytic result
on a $N_s^3\times 2N_s = 8^3 \times 16$ lattice (where $L = aN_s$)
is  $0.000277$, \cite{luscher96b}. Carrying out the interpolation
procedures outlined in the next section for a free configuration, with
background field given by eq.~(\ref{Uclassical}) yields $0.000271$.
For the stout smearing used here (see next section) we find
this is reduced to $\Delta M^{tree} = 0.000066$ and so we have
neglected $\Delta M^{tree}$ in the following and simply
used eq.~(\ref{improvement_condition}).


\section{The lattice simulation}
\label{lattice}


The $3$-flavour lattice simulation used the Chroma software library,
\cite{edwards04a}, the Schr\"odinger functional details following
\cite{klassen97a}. Results were mostly generated on
$N_s^3\times 2N_s \equiv 8^3\times 16$ lattices,
together with some additional $12^3\times 24$ lattices, using the
HMC algorithm together with the RHMC variation, \cite{clark06a},
for the $1$-flavour. A mild smearing of $\alpha = 0.1$ was used.
A series of simulations were performed (typically generating
$O(3000)$ trajectories for the $8^3\times 16$ lattices and
$O(2000)$ trajectories for the $12^3\times 24$ lattices), quadratic
and then linear interpolations of the $(M, \Delta M)$ results
being used to locate the optimal point, $(0, 0)$ as described below.
Some further details and tables of the results are given in
appendix~\ref{appendix_raw_results}. (Preliminary results were given
in \cite{cundy08a}.)


\subsection{$c_{sw}^*$}
\label{section_cswstar}

\begin{figure}[p]

\begin{minipage}{0.475\textwidth}

   \epsfxsize=7.00cm \epsfbox{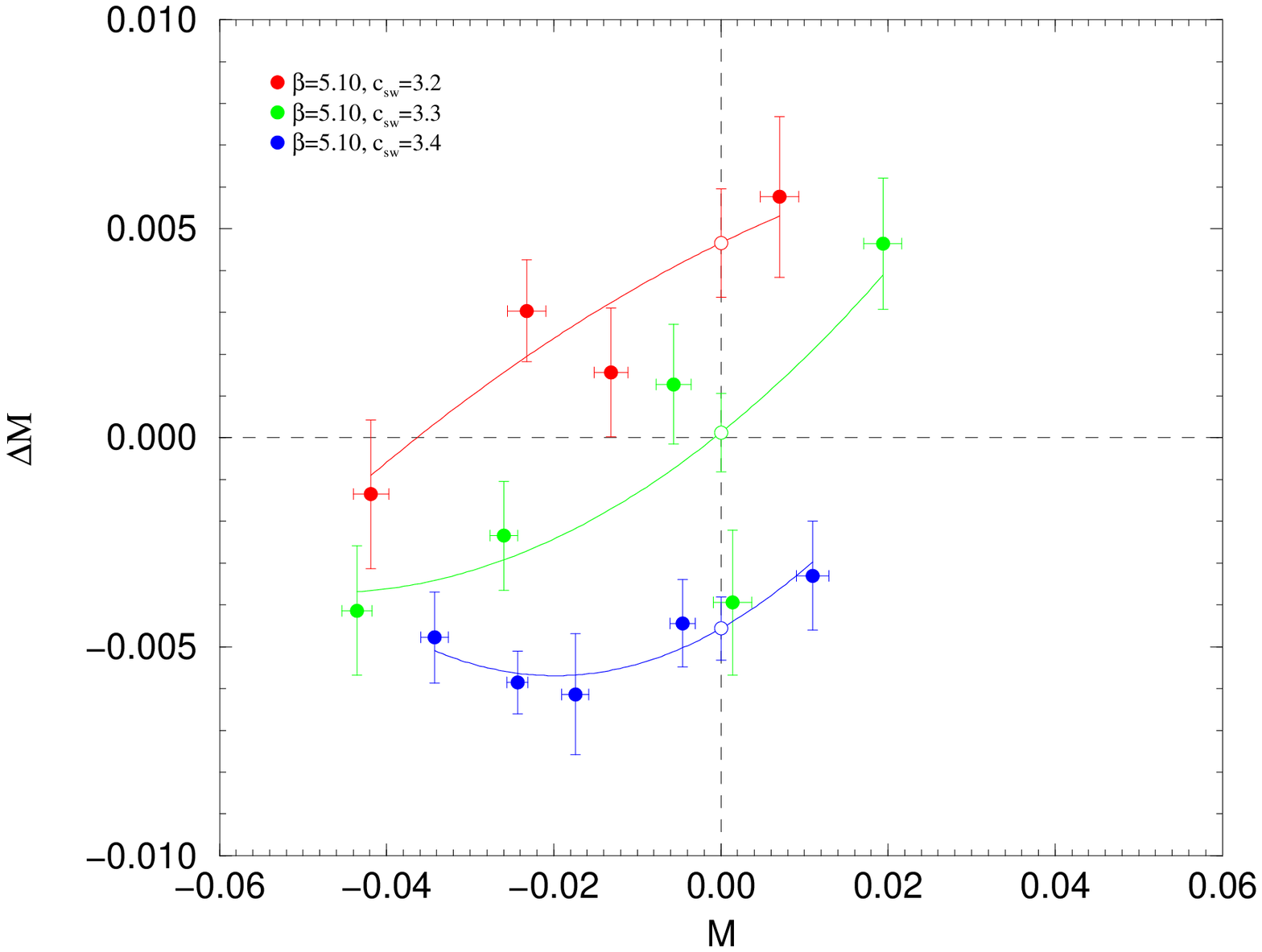}

\end{minipage} \hspace*{0.05\textwidth}
\begin{minipage}{0.475\textwidth}

   \epsfxsize=7.00cm \epsfbox{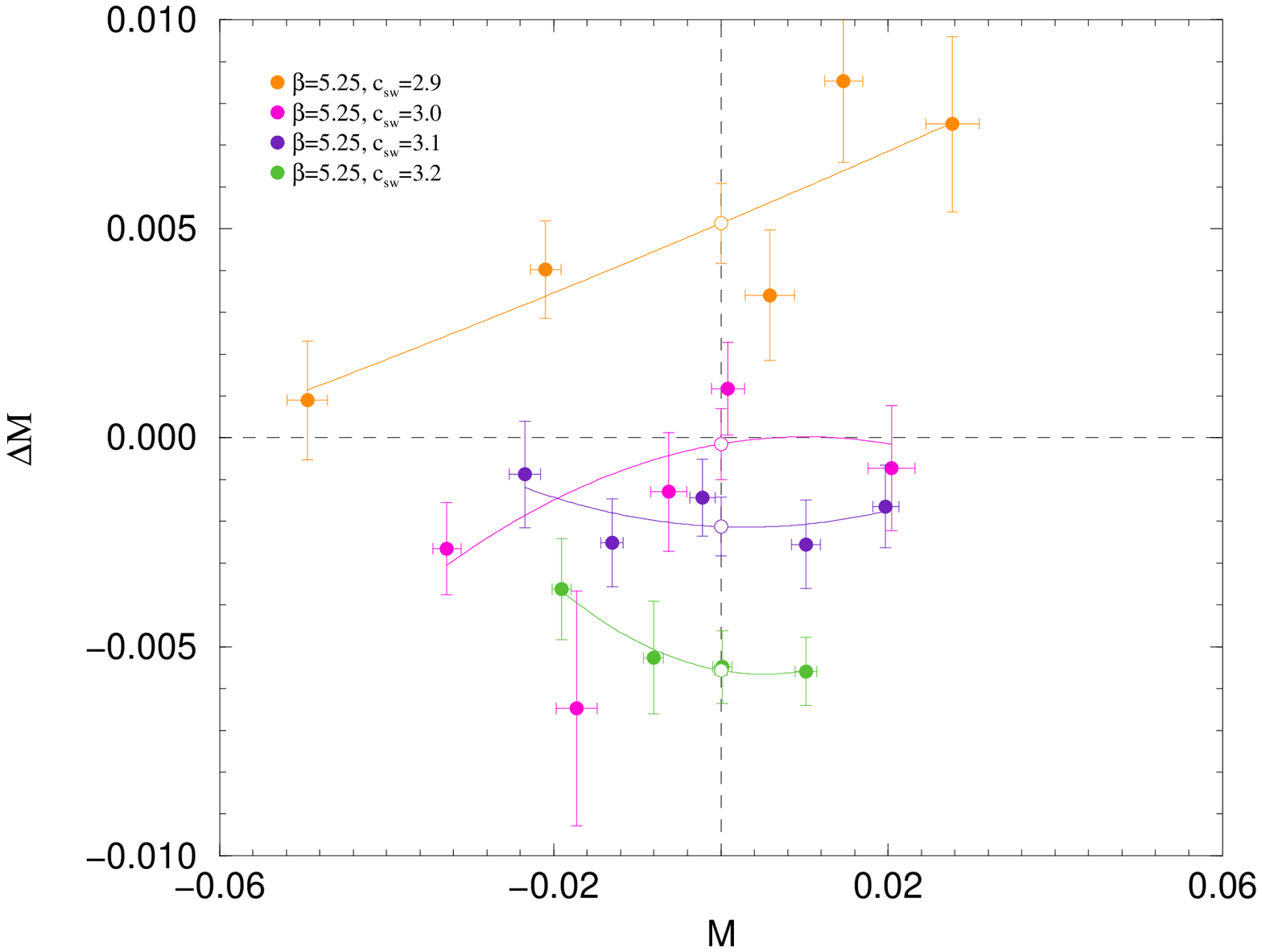}

\end{minipage}
\begin{minipage}{0.475\textwidth}

   \epsfxsize=7.00cm \epsfbox{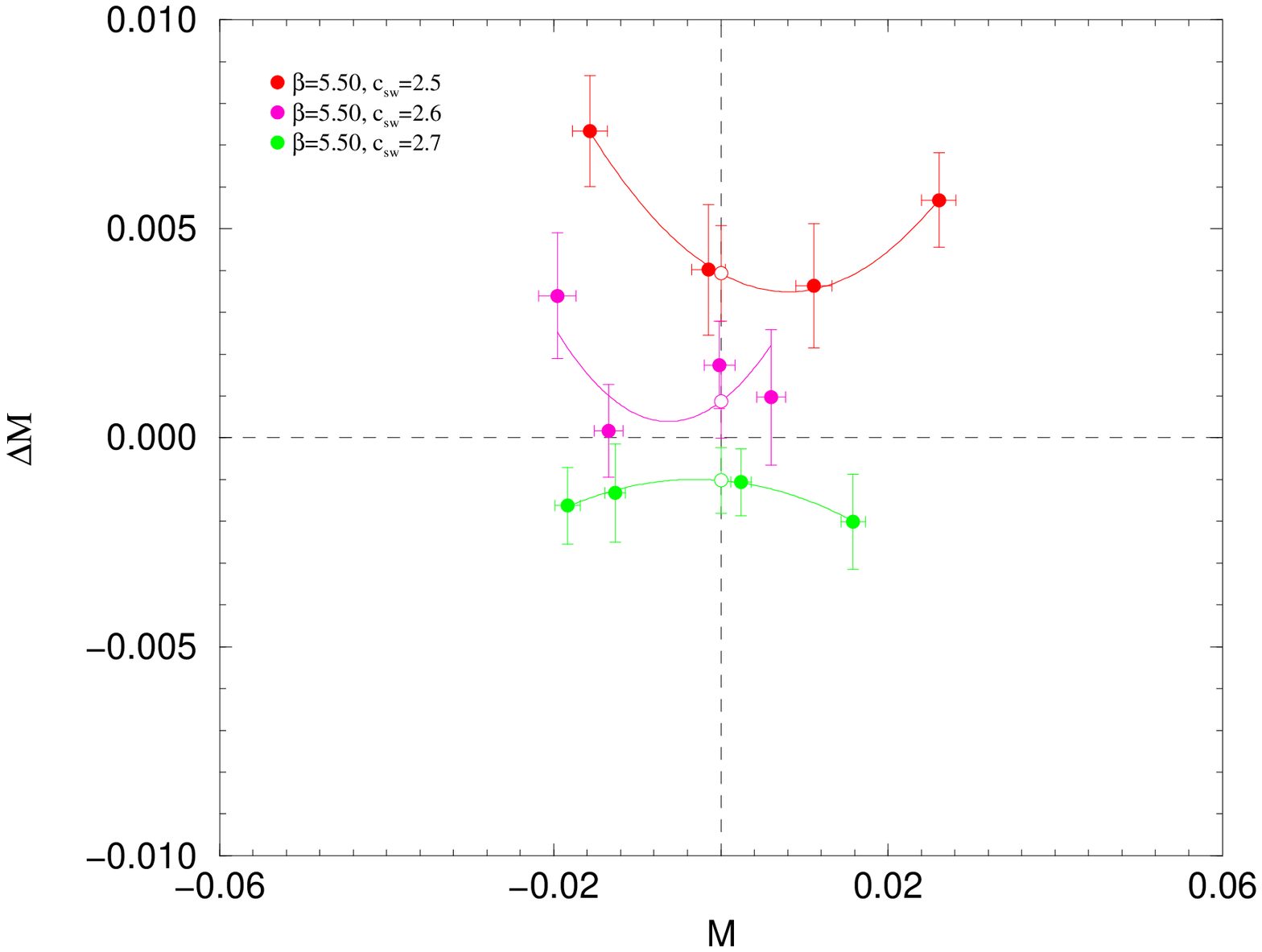}

\end{minipage} \hspace*{0.05\textwidth}
\begin{minipage}{0.475\textwidth}

   \epsfxsize=7.00cm \epsfbox{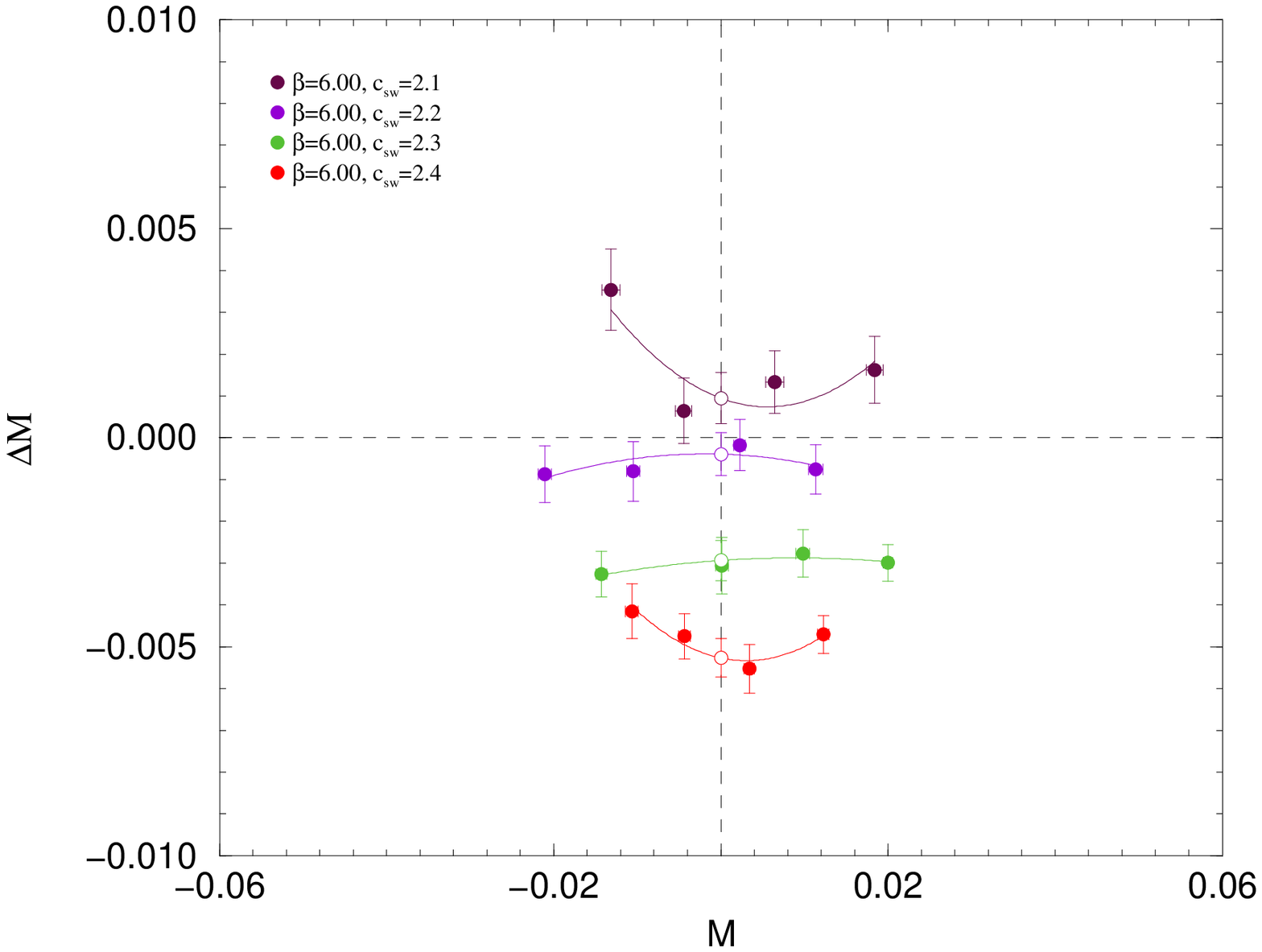}

\end{minipage}

\begin{minipage}{0.475\textwidth}

   \epsfxsize=7.00cm \epsfbox{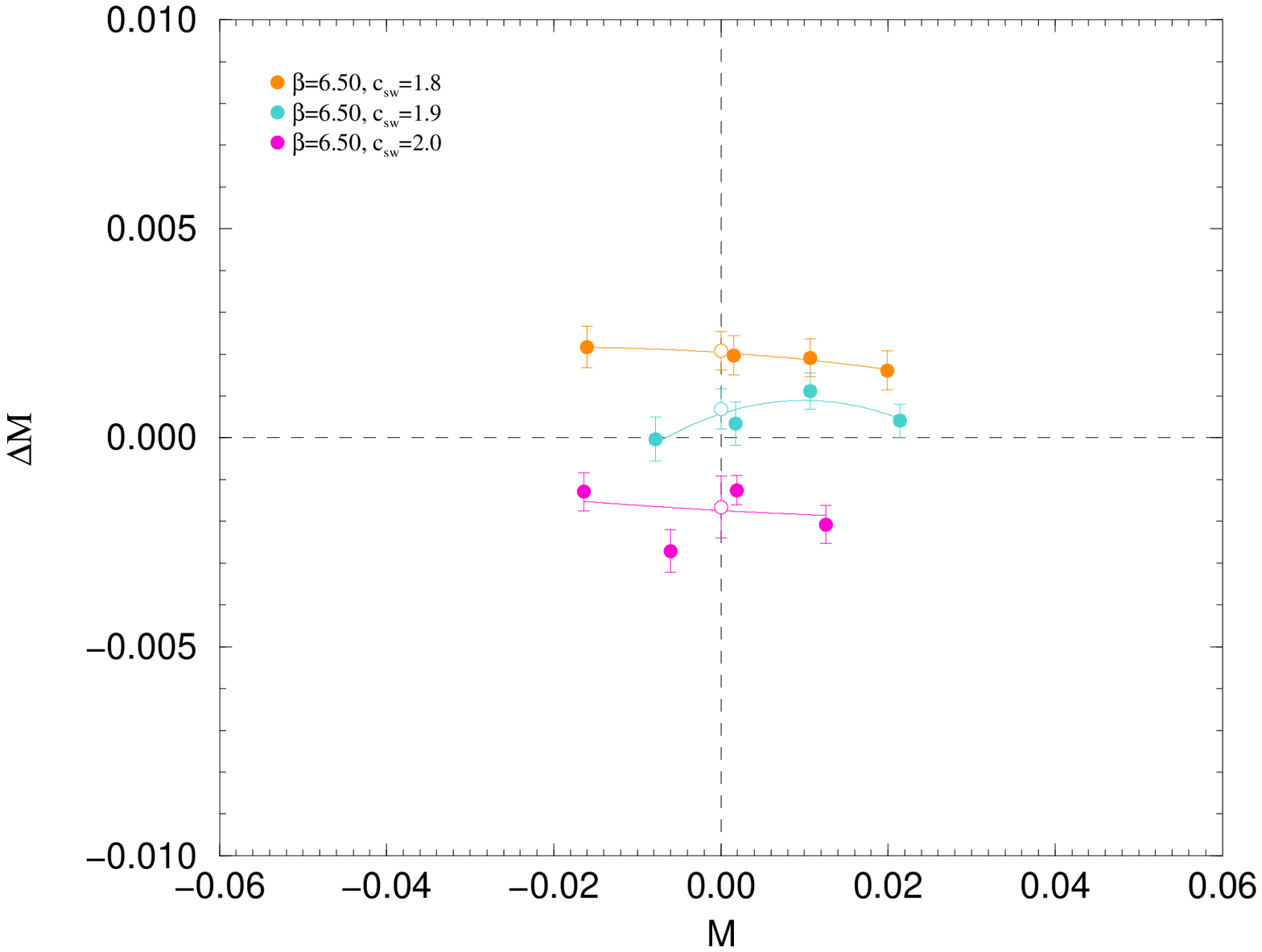}

\end{minipage} \hspace*{0.05\textwidth}
\begin{minipage}{0.475\textwidth}

   \epsfxsize=7.00cm \epsfbox{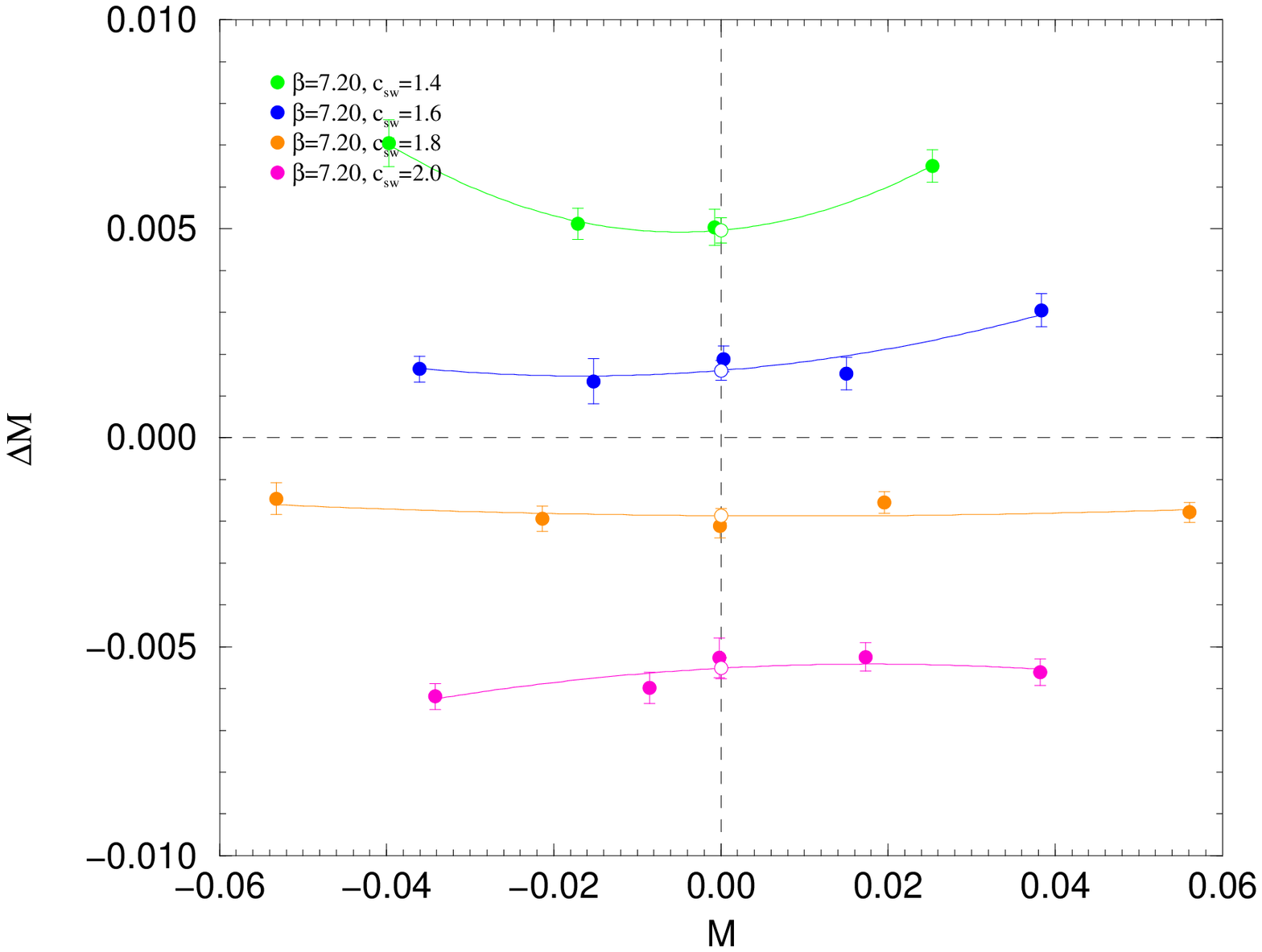}

\end{minipage}

\caption{$\Delta M$ against $M$ for $\beta = 5.10$, $5.25$
         (upper left, right pictures respectively)
         for $\beta = 5.50$, $\beta = 6.00$
         (middle left, right pictures respectively),
         and for $\beta = 6.50$, $\beta = 7.20$
         (lower left, right pictures respectively),
         together with quadratic interpolations to $M = 0$
         (the open symbols).}

\label{M-dM_b5p10-b6p00}

\end{figure}
\begin{figure}[p]

\begin{minipage}{0.475\textwidth}

   \epsfxsize=7.00cm \epsfbox{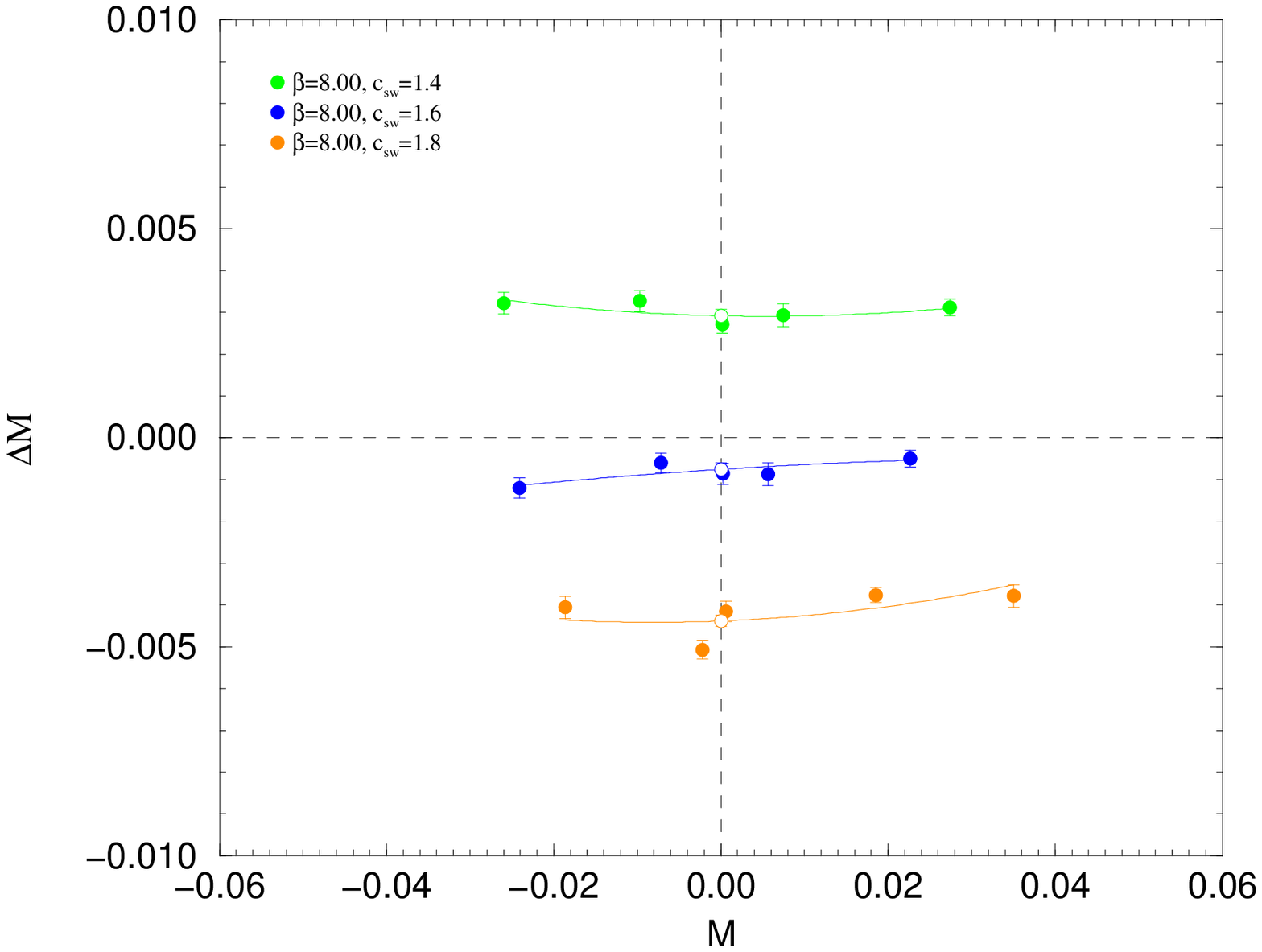}

\end{minipage} \hspace*{0.05\textwidth}
\begin{minipage}{0.475\textwidth}

   \epsfxsize=7.00cm \epsfbox{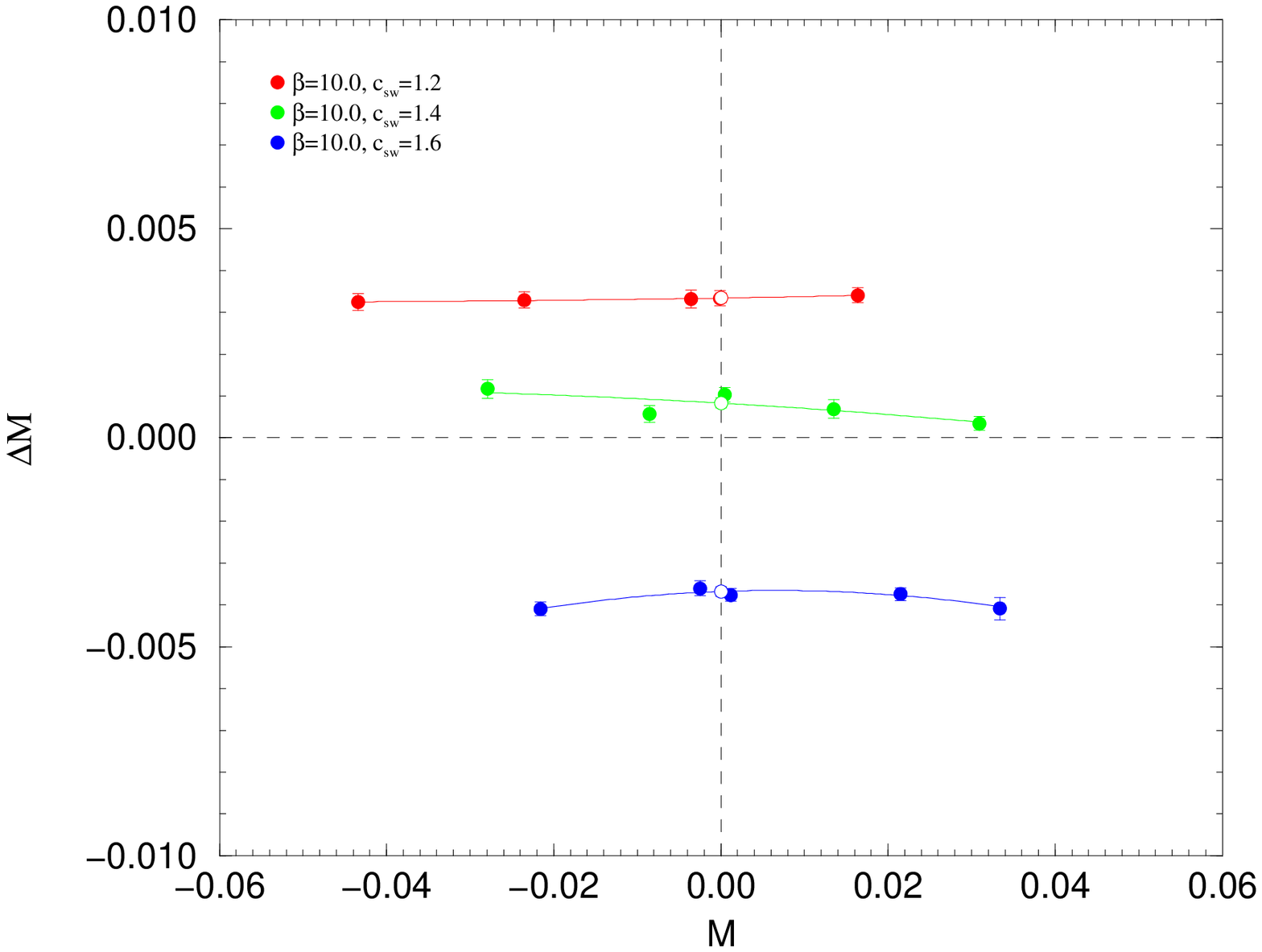}

\end{minipage}

\begin{minipage}{\textwidth}

   \hspace*{1.50in}
   \epsfxsize=7.00cm \epsfbox{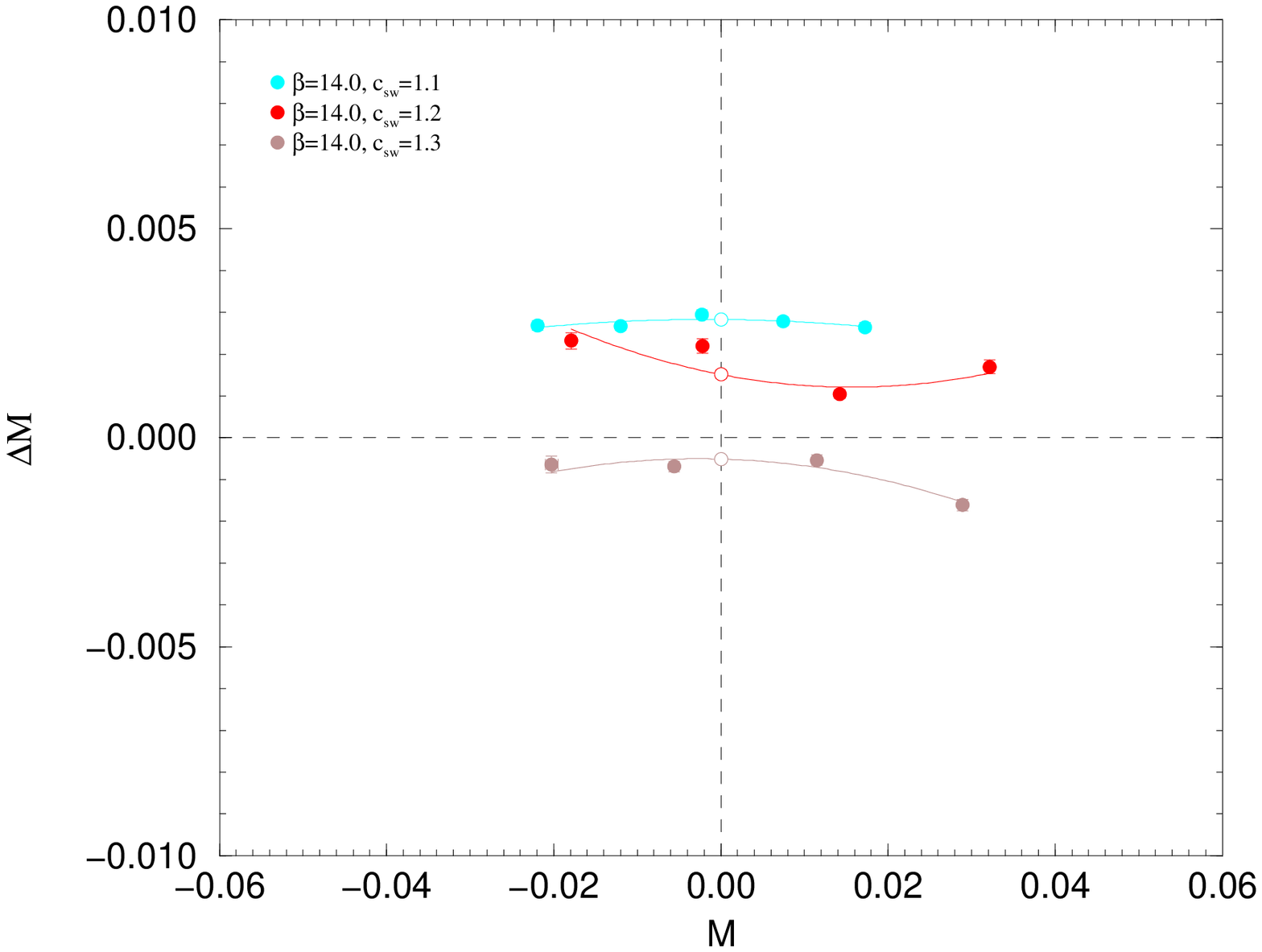}

\end{minipage}

\caption{$\Delta M$ against $M$ for $\beta = 8.00$, $10.0$
         (upper left, right pictures respectively)
         and for $\beta = 14.0$,
         (lower picture),
         together with quadratic interpolations to $M = 0$
         (the open symbols).}

\label{M-dM_b8p00-b14p0}

\end{figure}

We have a two-parameter interpolation in $c_{sw}$ and $\kappa$
which is split here into two separate interpolations.
First plotting $\Delta M$ against $M$ and then interpolating to $M = 0$
for fixed $c_{sw}$ gives a critical $\kappa$ namely $\kappa_c(c_{sw})$,
\begin{equation}
   \Delta M(c_{sw},\kappa)|_{M=0} 
      \equiv \Delta M(c_{sw},\kappa_c(c_{sw}))|_{M=0} 
      \equiv \Delta M(c_{sw}) \,.
\end{equation}
In Figs.~\ref{M-dM_b5p10-b6p00}, \ref{M-dM_b8p00-b14p0}
\begin{figure}[p]

\begin{minipage}{0.475\textwidth}

   \epsfxsize=7.00cm \epsfbox{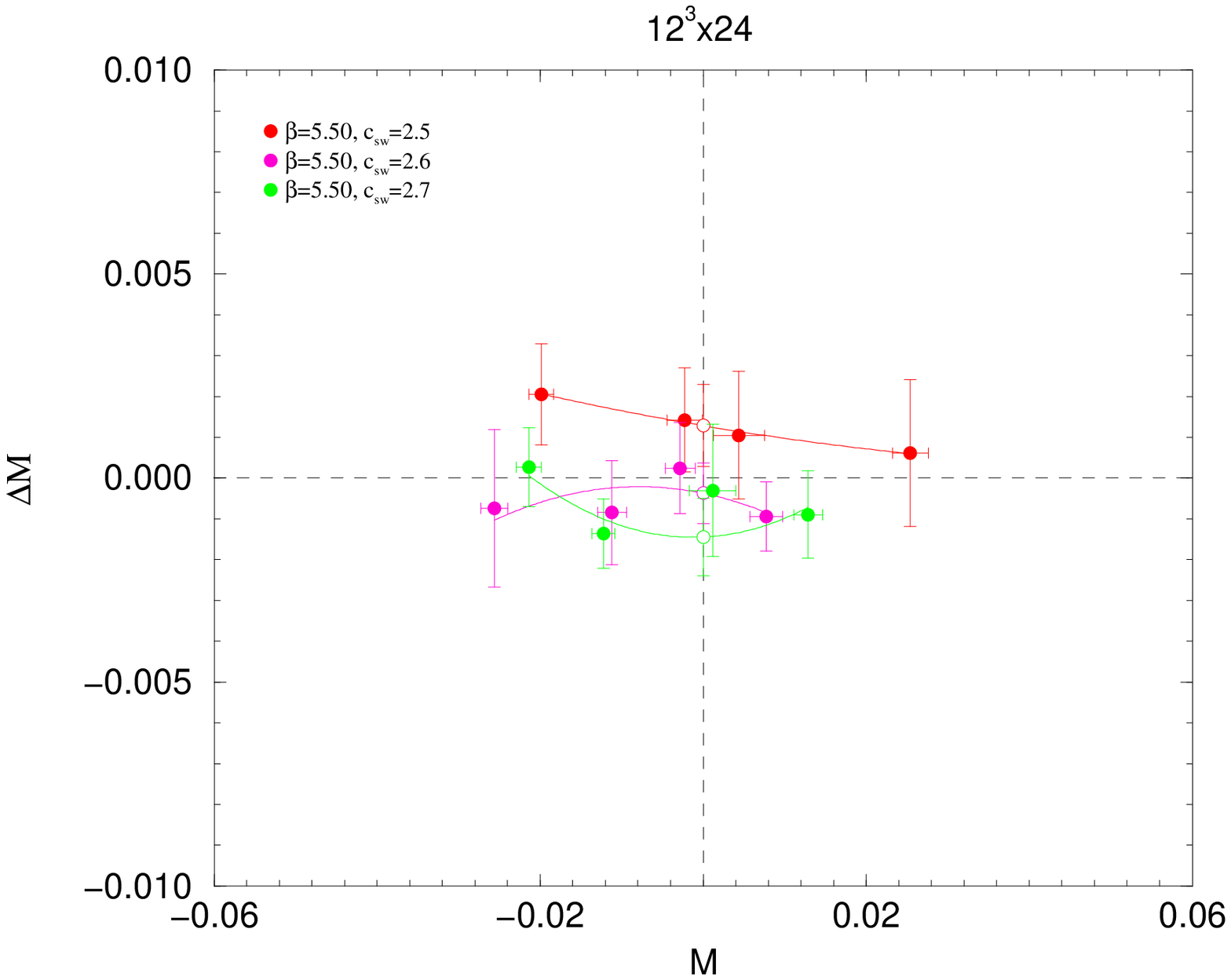}

\end{minipage} \hspace*{0.05\textwidth}
\begin{minipage}{0.475\textwidth}

   \epsfxsize=7.00cm \epsfbox{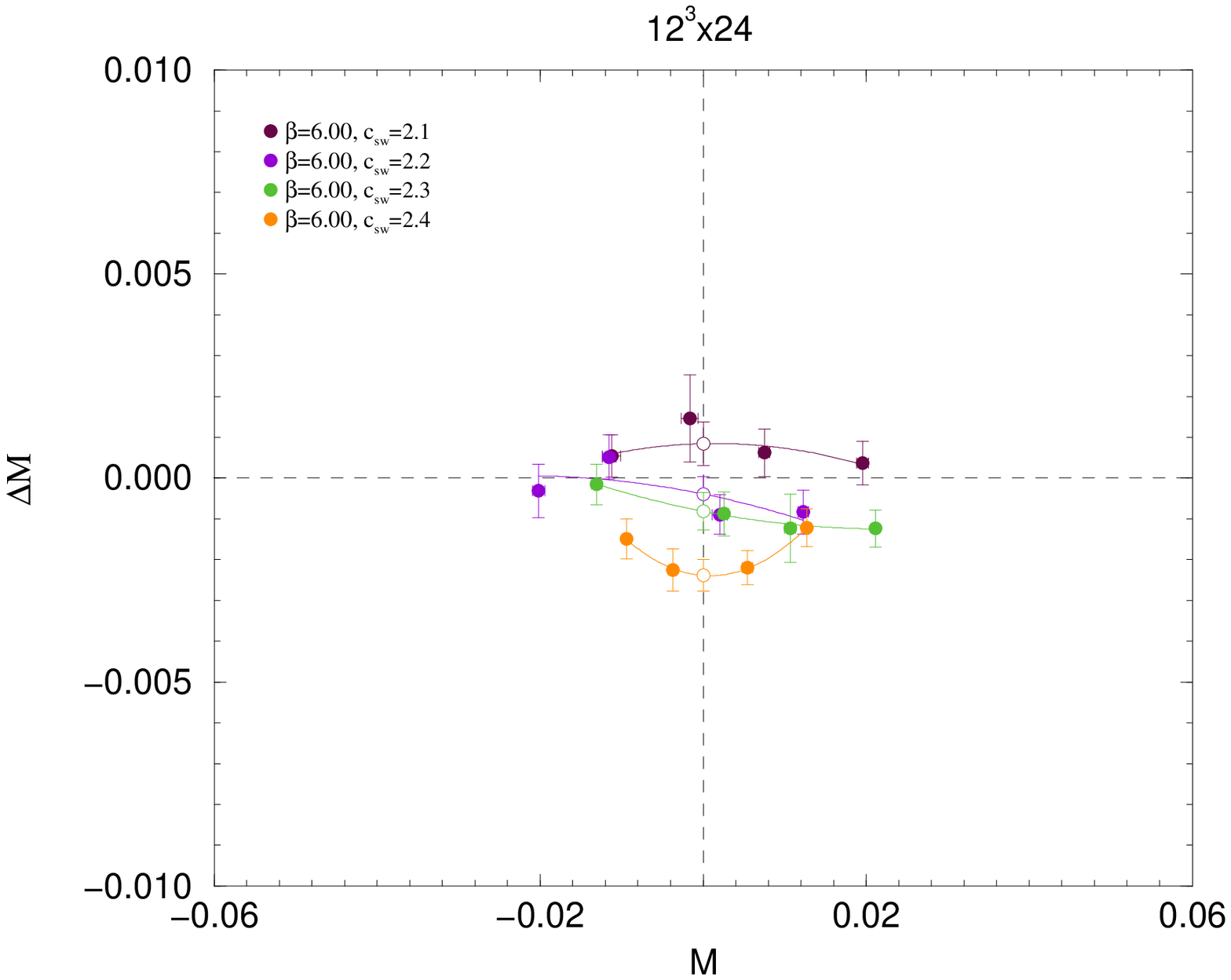}

\end{minipage}

\caption{$\Delta M$ against $M$ for $\beta = 5.50$, $6.00$
         (left, right pictures respectively) on a $12^3\times 24$
         lattice together with quadratic interpolations to $M = 0$
         (the open symbols).}

\label{M-dM_b5p50-b6p00_12x24}

\end{figure}
we plot $\Delta M$ versus $M$ for various $c_{sw}$ values
for the $8^3\times 16$ lattices and in Fig.~\ref{M-dM_b5p50-b6p00_12x24}
the results for the $12^3 \times 24$ lattices.

These graphs are the fundamental plots requiring high
statistics as $\Delta M$ is the difference between two
different $M$s. As there are always $4$ (or more) points
for each graph a quadratic fit is made and the value of
$\Delta M$ is determined where $M$ vanishes.

\begin{figure}[t]
\begin{minipage}{0.475\textwidth}

   \epsfxsize=7.00cm \epsfbox{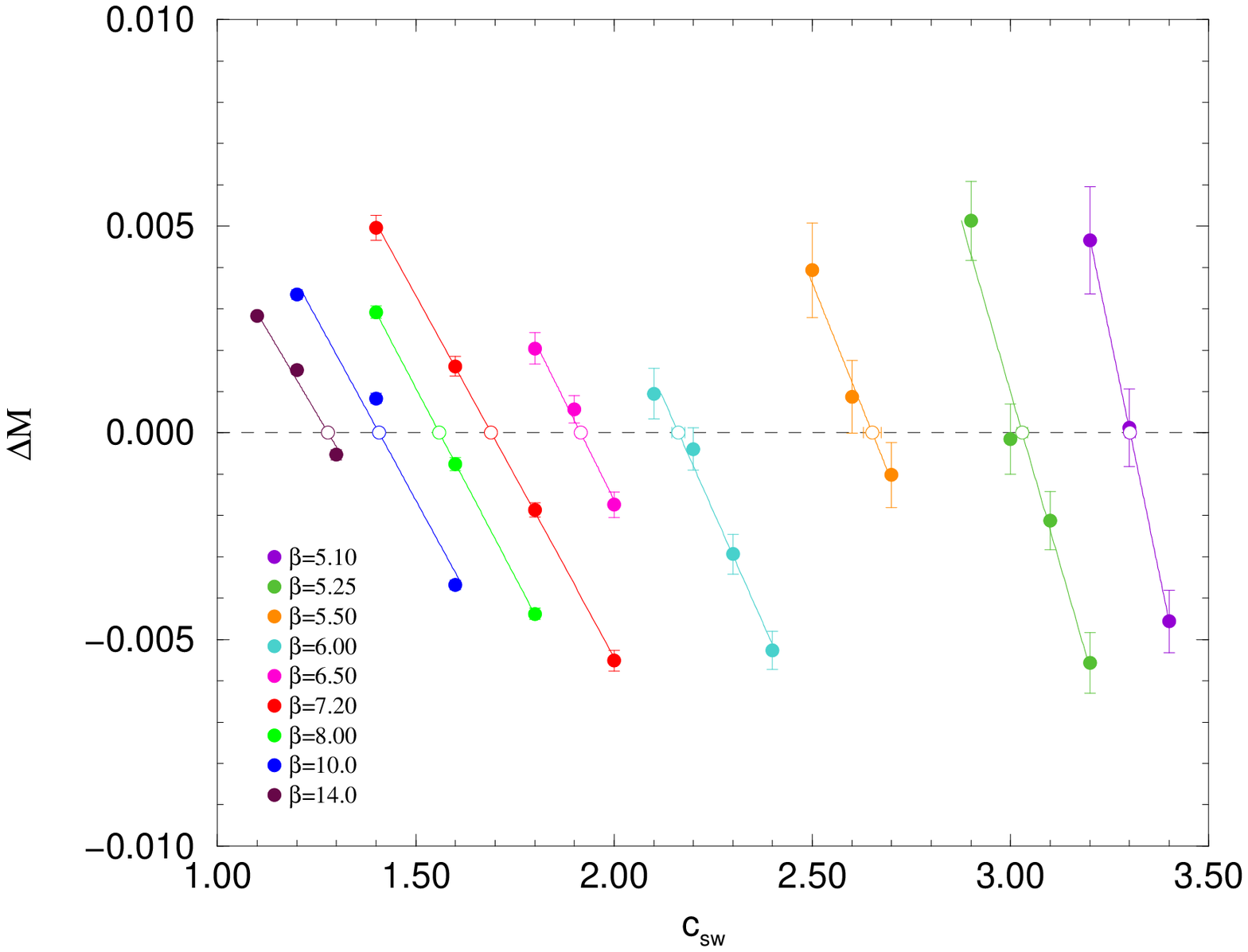}

\end{minipage} \hspace*{0.05\textwidth}
\begin{minipage}{0.475\textwidth}

   \epsfxsize=7.00cm \epsfbox{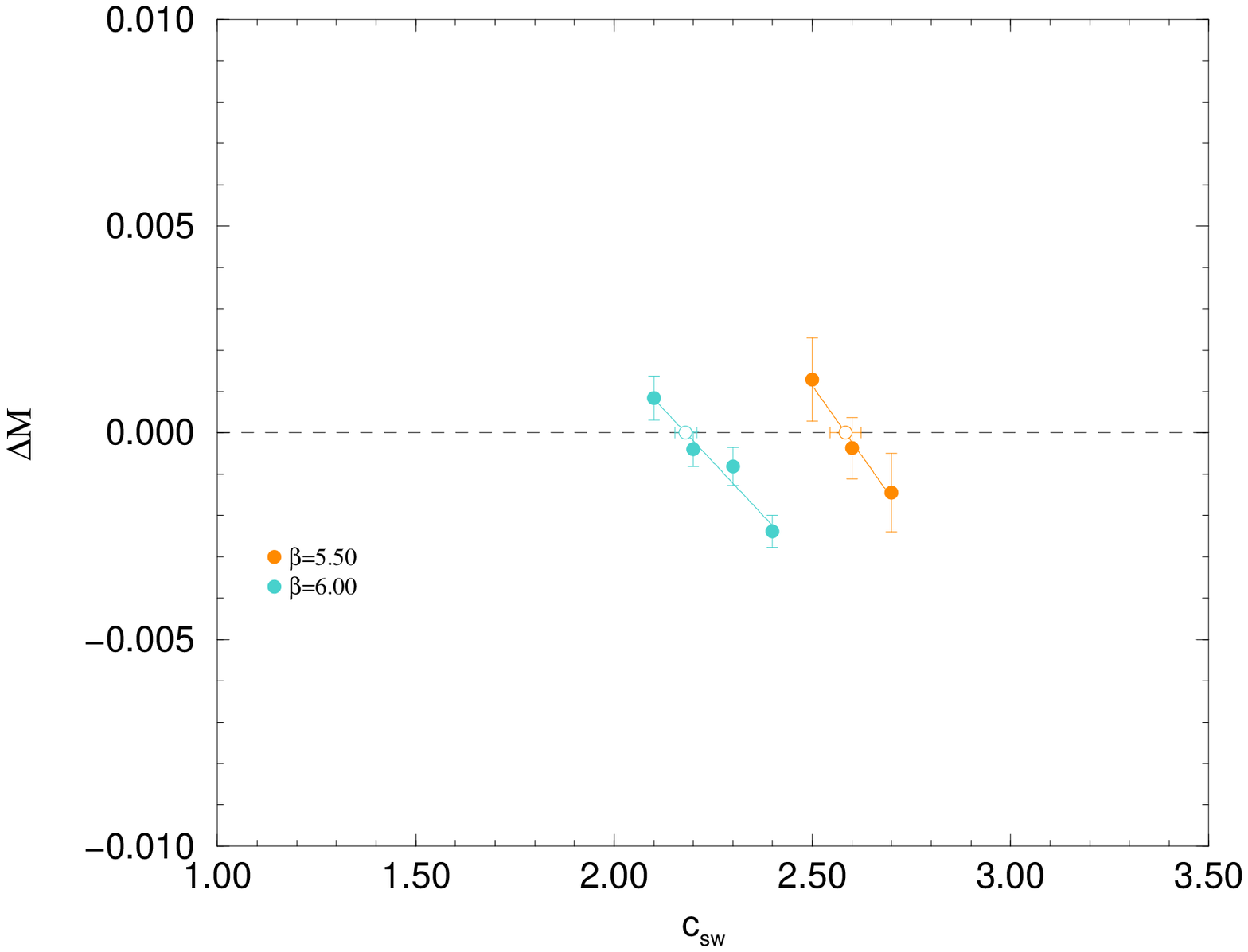}

\end{minipage}

\caption{$\Delta M$ at $M = 0$ against $c_{sw}$ for various values
         of $\beta$ (filled circles) together with linear interpolations
         to $\Delta M = 0$ (open circles). The left plot shows the
         $8^3\times 16$ results while the right plot shows the
         $12^3\times 24$ results.}

\label{csw_dM}

\end{figure}

These values of $\Delta M(c_{sw})$ for each $\beta$
value are then plotted against $c_{sw}$ as shown in 
Fig.~\ref{csw_dM}
together with linear fits. The point where $\Delta M(c_{sw})$ vanishes
gives $c_{sw}^*$. This gives values of
\begin{equation}
   c_{sw}^* = \left\{ 
      \begin{array}{l}
          \left. \begin{array}{lc}
                    3.302(13) & \beta = 5.10 \\
                    3.030(13) & \beta = 5.25 \\
                    2.651(23) & \beta = 5.50 \\
                    2.163(17) & \beta = 6.00 \\
                    1.915(10) & \beta = 6.50 \\
                    1.690(07) & \beta = 7.20 \\
                    1.559(05) & \beta = 8.00 \\
                    1.407(04) & \beta = 10.0 \\
                    1.279(06) & \beta = 14.0 \\
                 \end{array}
          \right\} \, 8^3\times 16 \\
                                   \\
          \left. \begin{array}{lc}
                    2.584(38) & \beta = 5.50 \\
                    2.181(28) & \beta = 6.00 \\
                 \end{array}
          \right\} \, 12^3\times 24 \\
      \end{array}  
               \right.
\label{cswstar_vals}
\end{equation}
We postpone a discussion of possible finite size effects until
section~\ref{finite_size_effects}.

From Fig.~\ref{csw_dM}, we see that linear fits even for
four points (the $\beta = 7.20$, $6.00$, $5.25$ results) show very
little curvature, so that we may write, \cite{jansen98a}
\begin{equation}
   \Delta M(c_{sw}) = \omega\,(c_{sw} - c_{sw}^*) \,,
\label{deltaM_csw_linear_fit}
\end{equation}
with the gradient, $\omega$, a slowly varying function of $g_0$.
To test this we note that
\begin{equation}
   {\partial\Delta M(c_{sw}) \over \partial c_{sw}} = \omega \,,
\end{equation}
so a fit to the gradients in Fig.~\ref{csw_dM} (for the $8^3\times 16$
lattices) yields an estimate for $\omega$. We find that $\omega$
is constant with an approximate value of $-0.018$, although for the
largest values of $g_0^2$ there are deviations from this.


\subsection{$\kappa_c^*$}


A similar procedure yields $\kappa_c^*$: plotting $M$ against $1/\kappa$
and interpolating quadratically to $M = 0$ for fixed $c_{sw}$ gives the
critical $\kappa$, denoted by $\kappa_c(c_{sw})$. Then subsequently plotting
$\Delta M(c_{sw})$ against $1/\kappa_c(c_{sw})$ and interpolating
using a linear fit to $\Delta M = 0$ gives $\kappa_c^*$.

\begin{figure}[p]

\begin{minipage}{0.475\textwidth}

   \epsfxsize=7.00cm \epsfbox{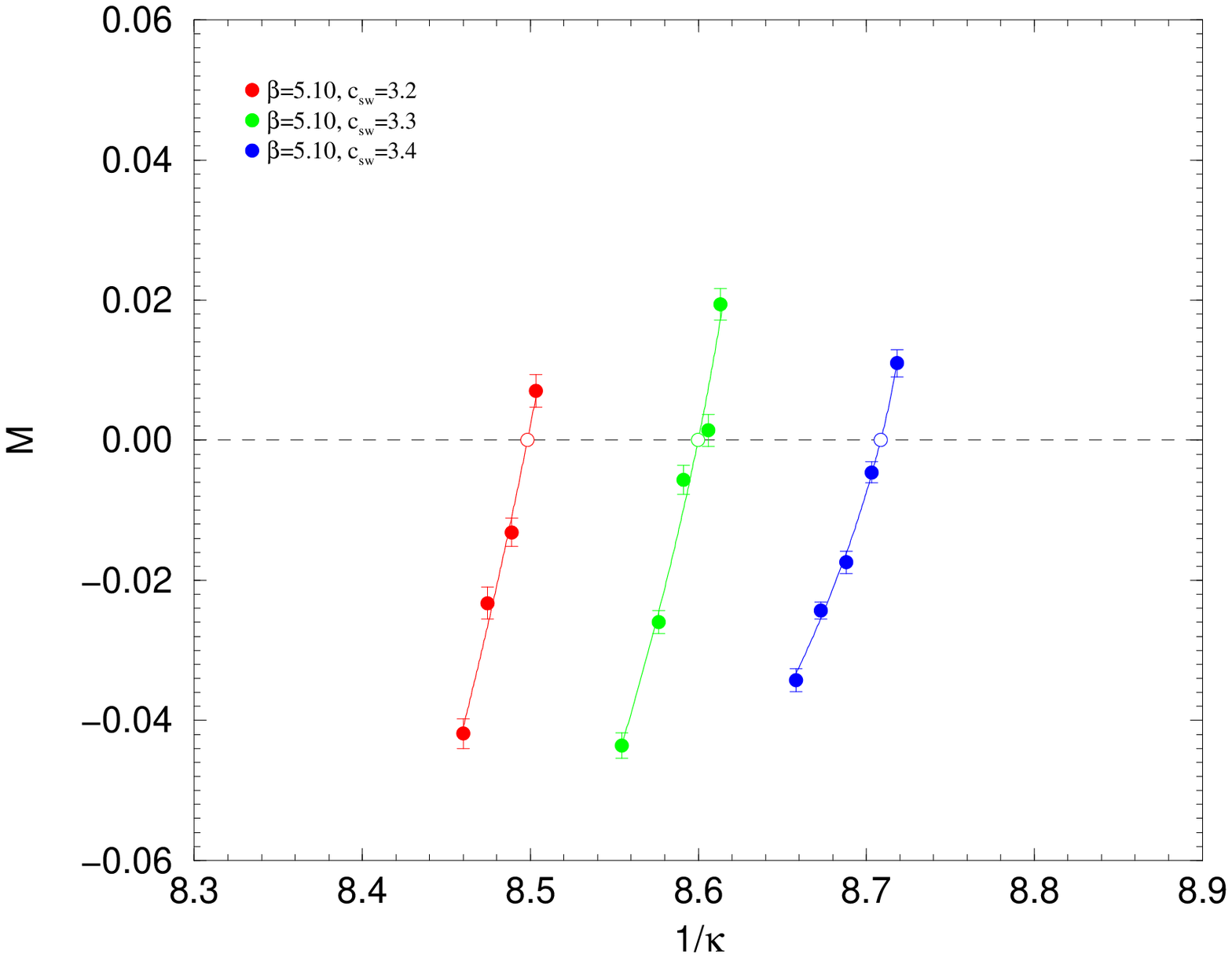}

\end{minipage} \hspace*{0.05\textwidth}
\begin{minipage}{0.475\textwidth}

   \epsfxsize=7.00cm \epsfbox{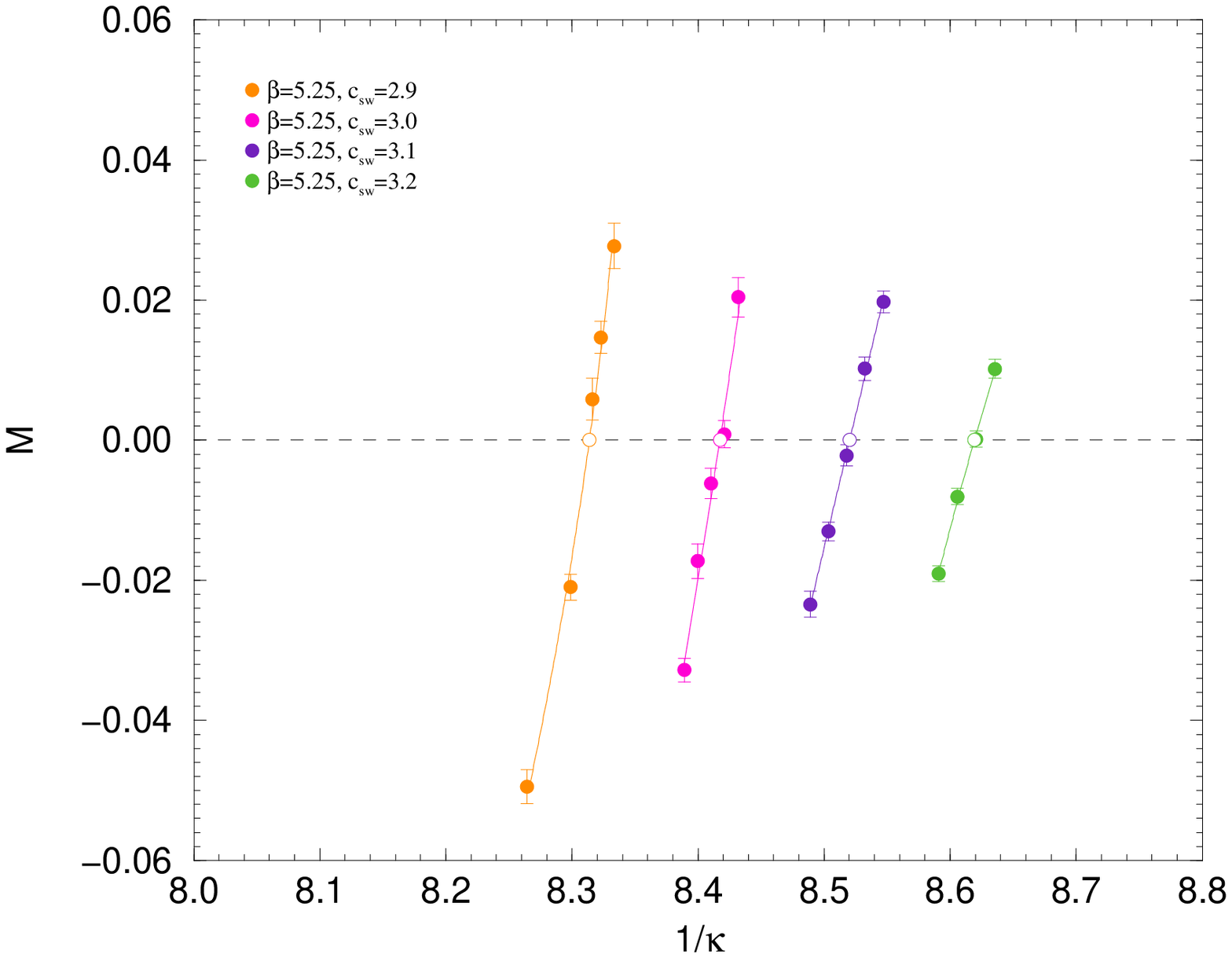}

\end{minipage}

\begin{minipage}{0.475\textwidth}

   \epsfxsize=7.00cm \epsfbox{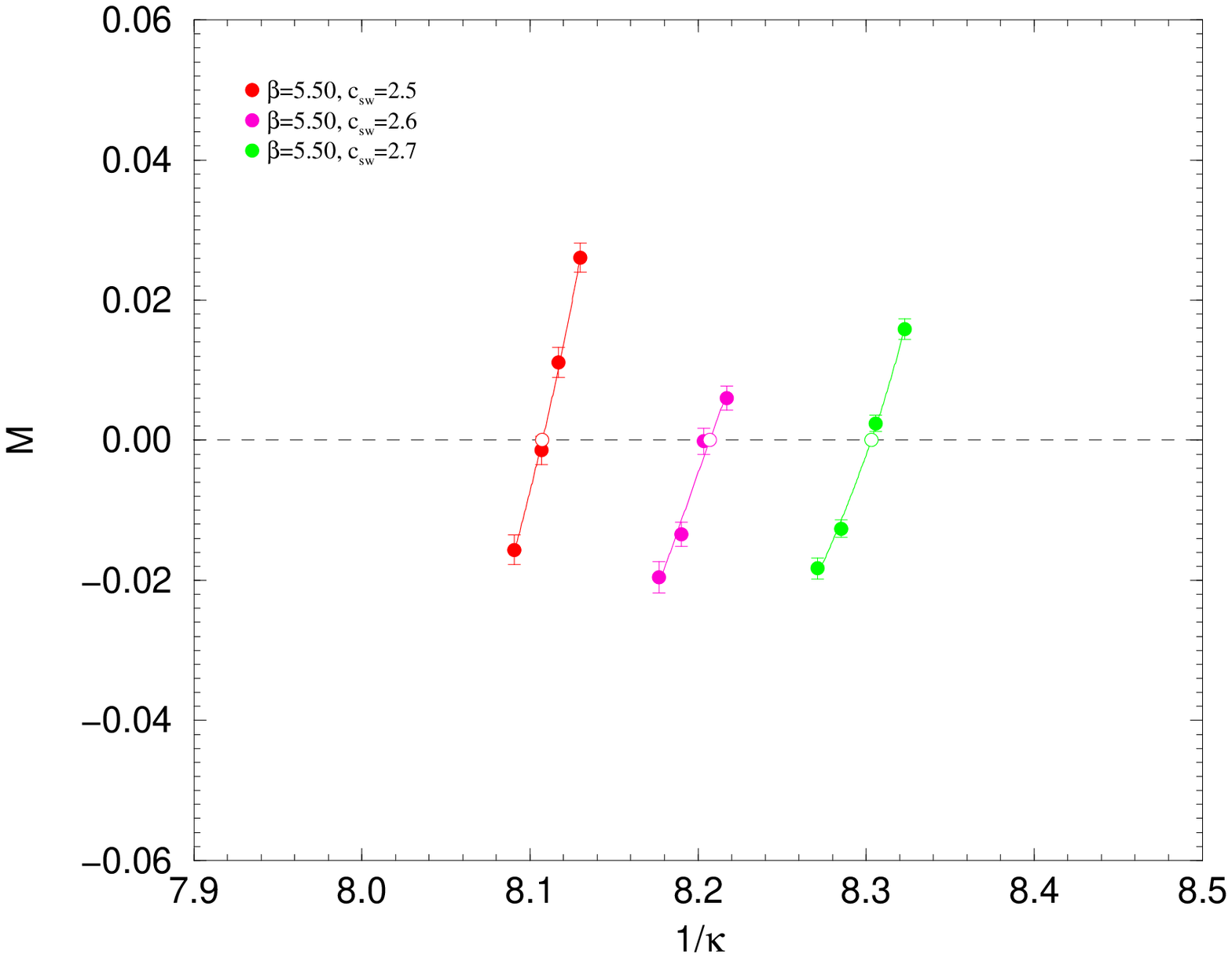}

\end{minipage} \hspace*{0.05\textwidth}
\begin{minipage}{0.475\textwidth}

   \epsfxsize=7.00cm \epsfbox{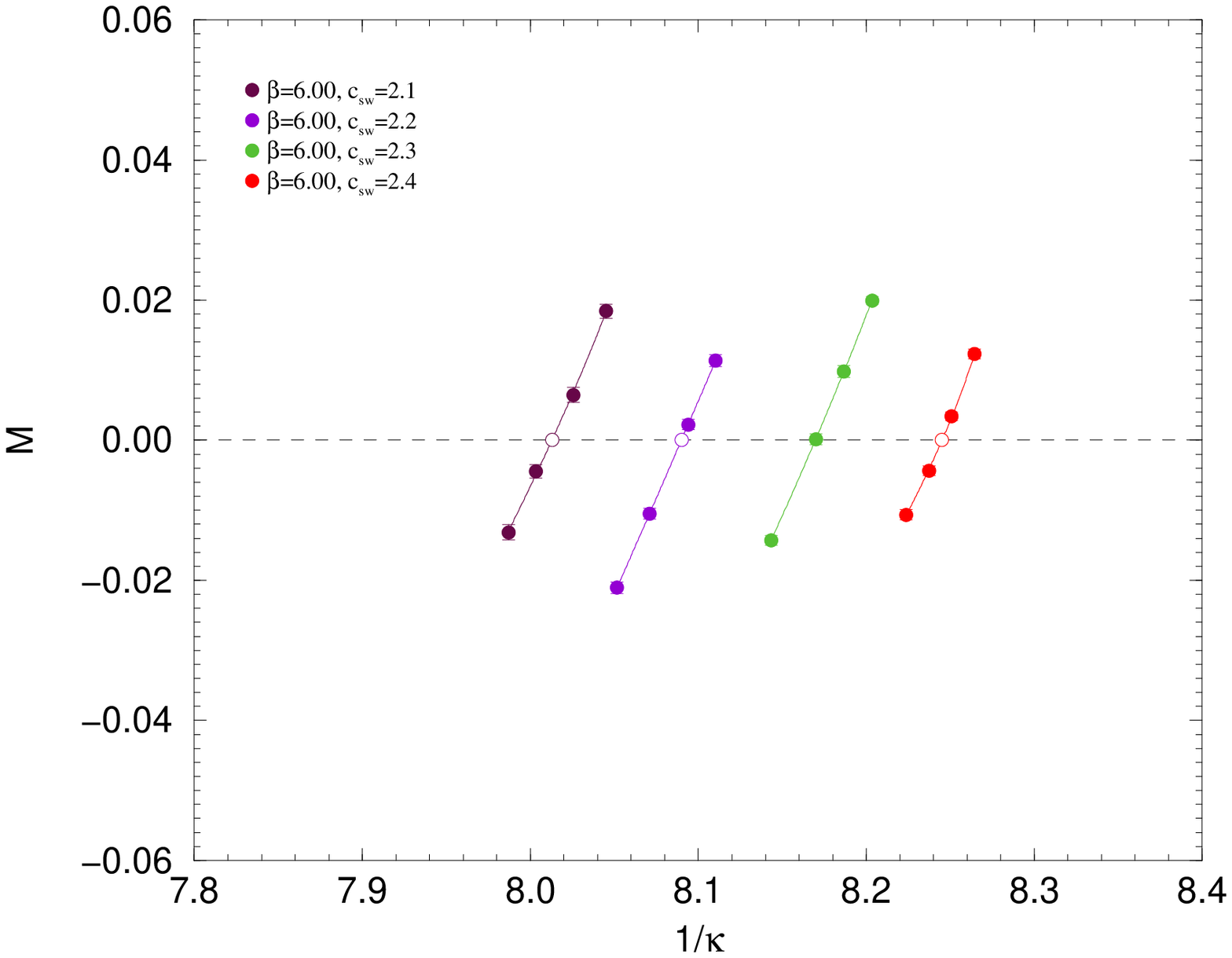}

\end{minipage}

\begin{minipage}{0.475\textwidth}

   \epsfxsize=7.00cm \epsfbox{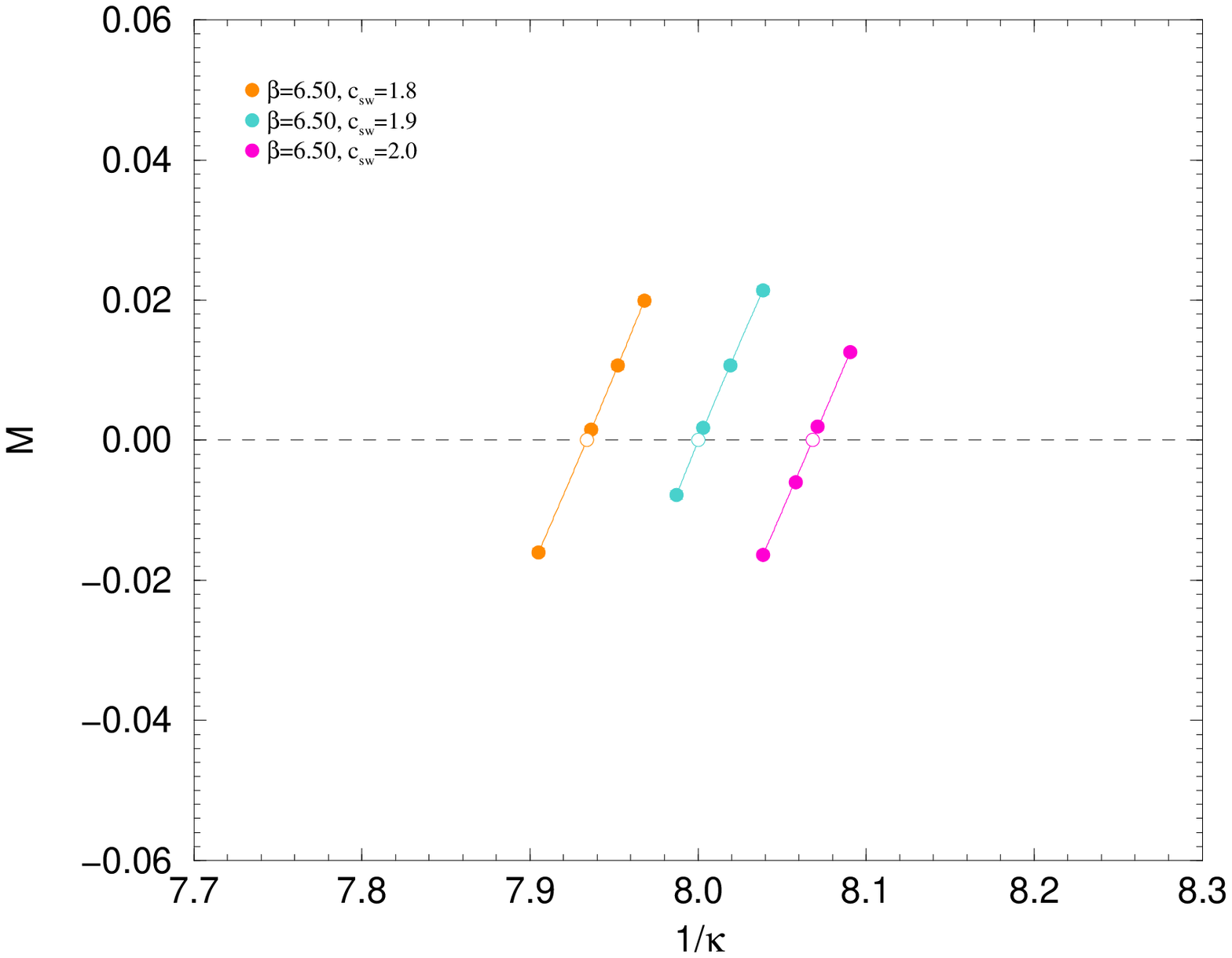}

\end{minipage} \hspace*{0.05\textwidth}
\begin{minipage}{0.475\textwidth}

   \epsfxsize=7.00cm \epsfbox{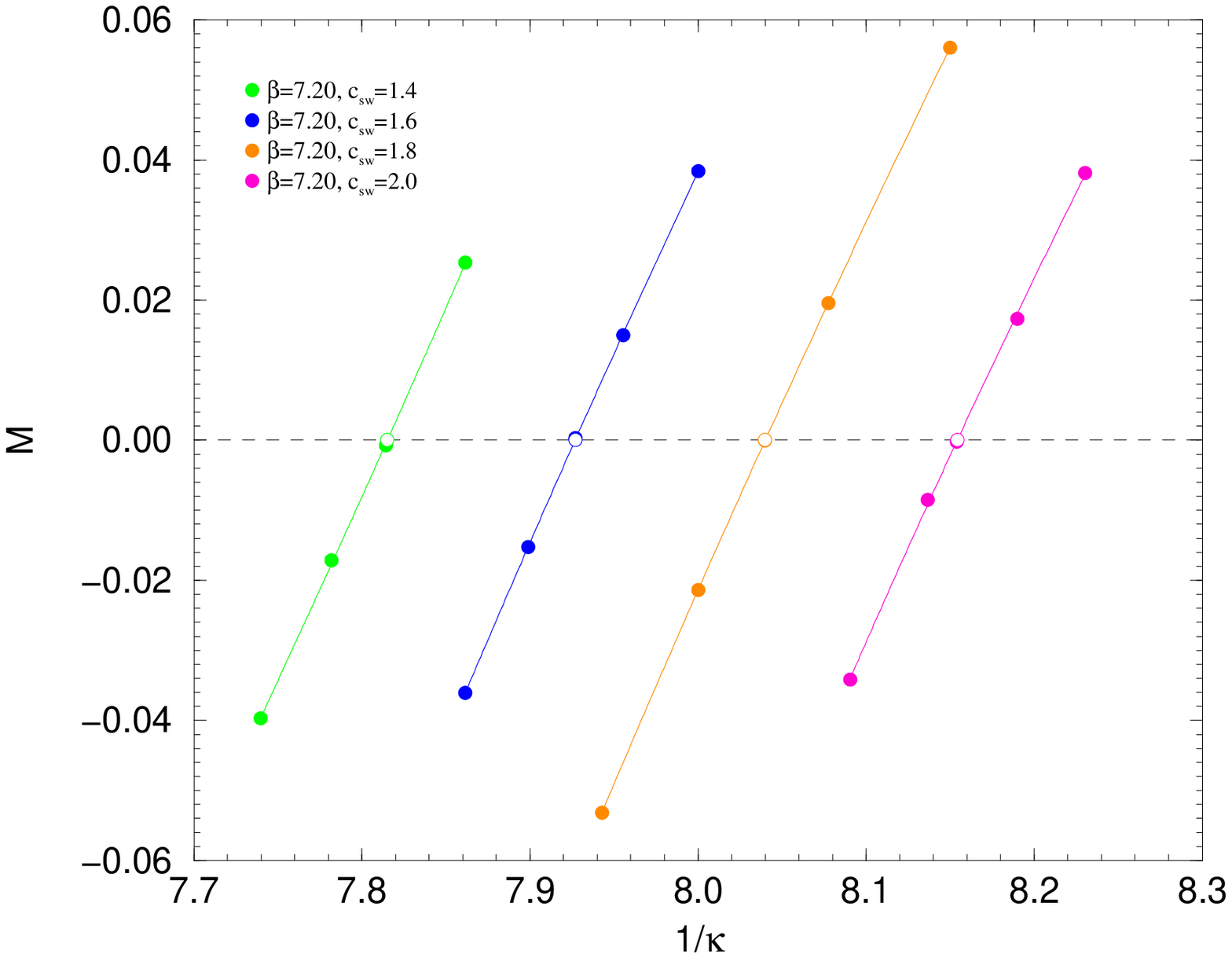}

\end{minipage}

\caption{$M$ against $1/\kappa$ for $\beta = 5.10$, $5.25$
         (upper left, right pictures respectively)
         for $\beta = 5.50$, $\beta = 6.00$
         (middle left, right pictures respectively),
         and for $\beta = 6.50$, $\beta = 7.20$
         (lower left, right pictures respectively),
         together with quadratic interpolations to $M = 0$
         (the open symbols).}

\label{ookap_M_b5p10-b6p00}

\end{figure}

\begin{figure}[p]

\begin{minipage}{0.475\textwidth}

   \epsfxsize=7.00cm \epsfbox{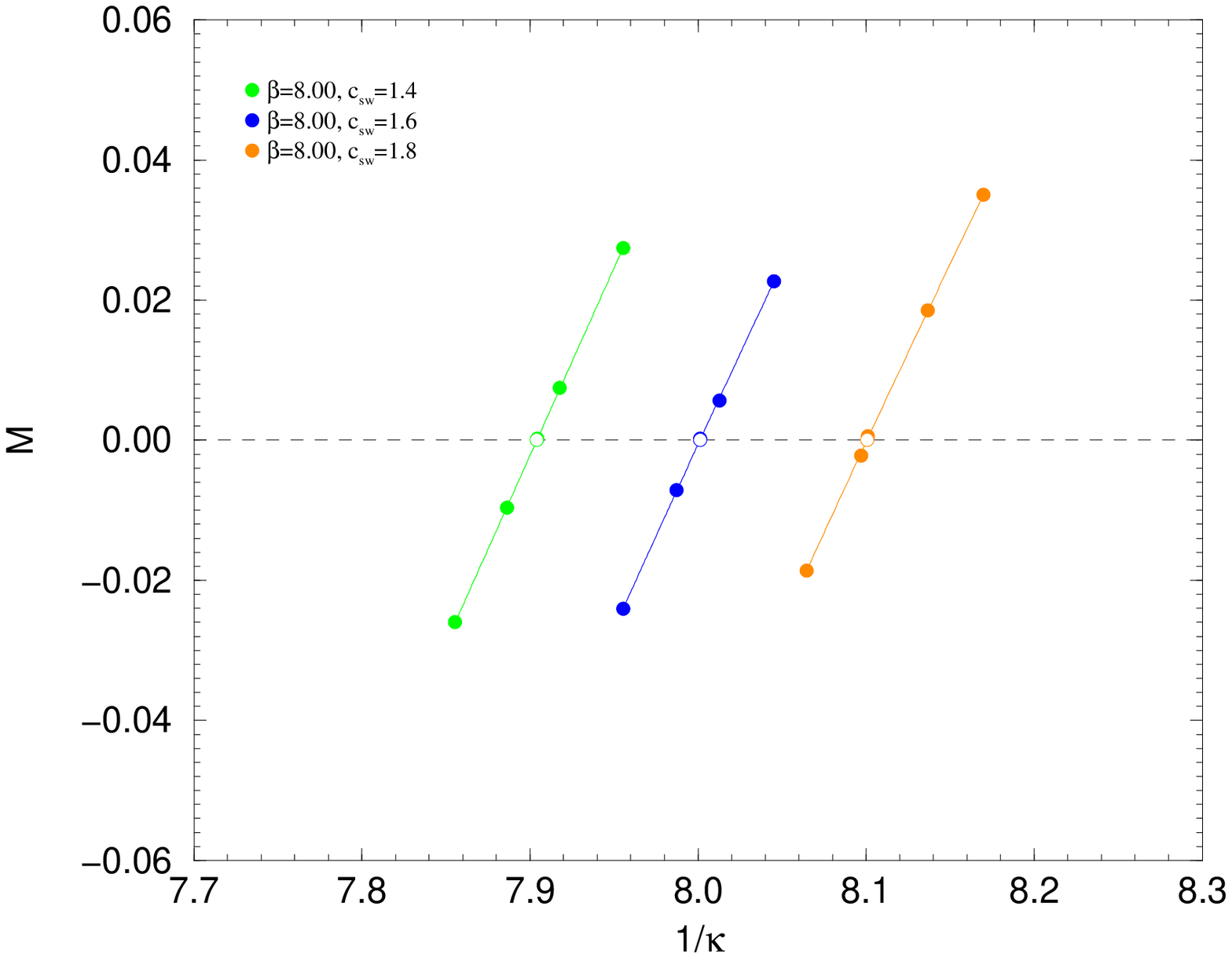}

\end{minipage} \hspace*{0.05\textwidth}
\begin{minipage}{0.475\textwidth}

   \epsfxsize=7.00cm \epsfbox{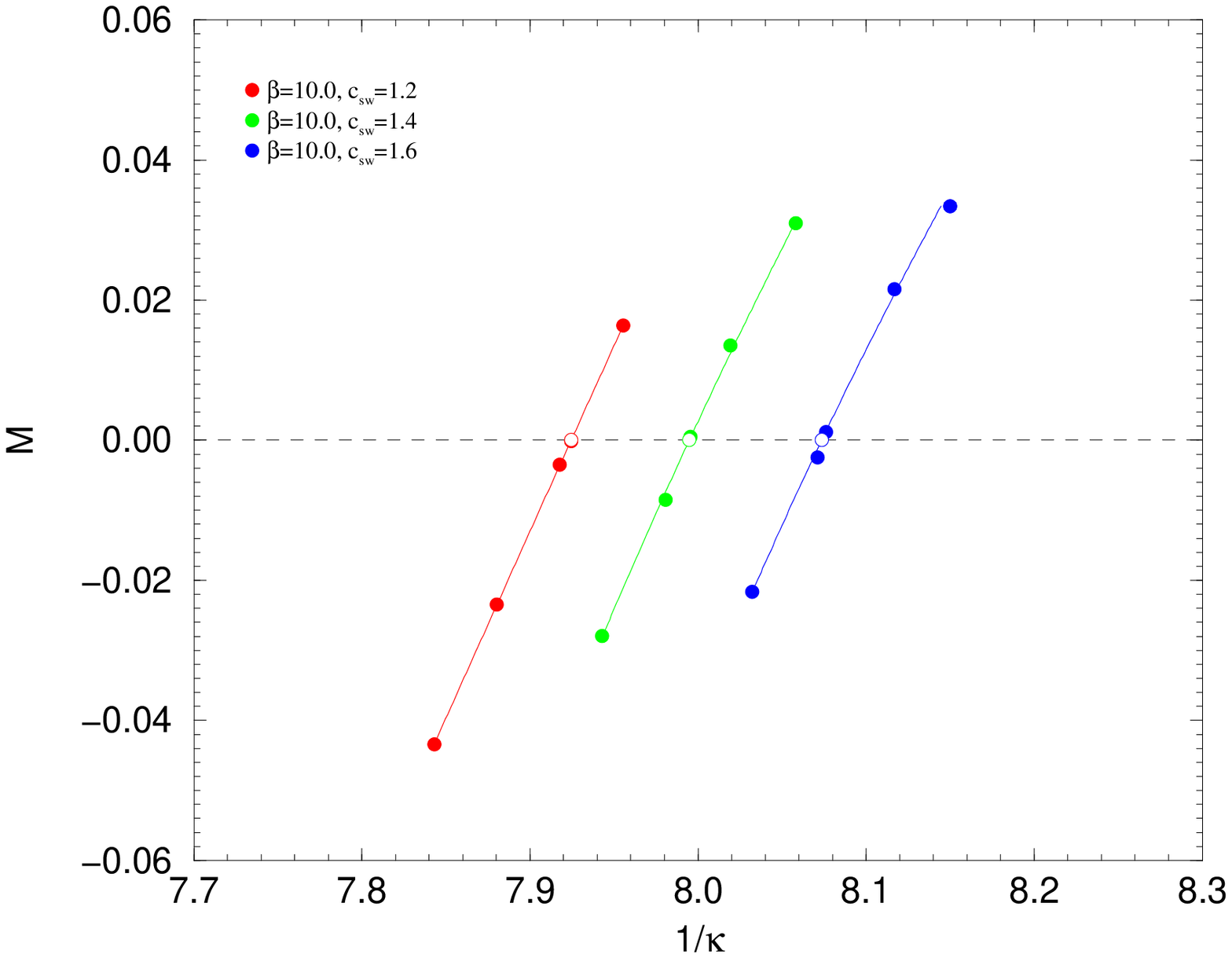}

\end{minipage}

\begin{minipage}{\textwidth}

   \hspace*{1.50in}
   \epsfxsize=7.00cm \epsfbox{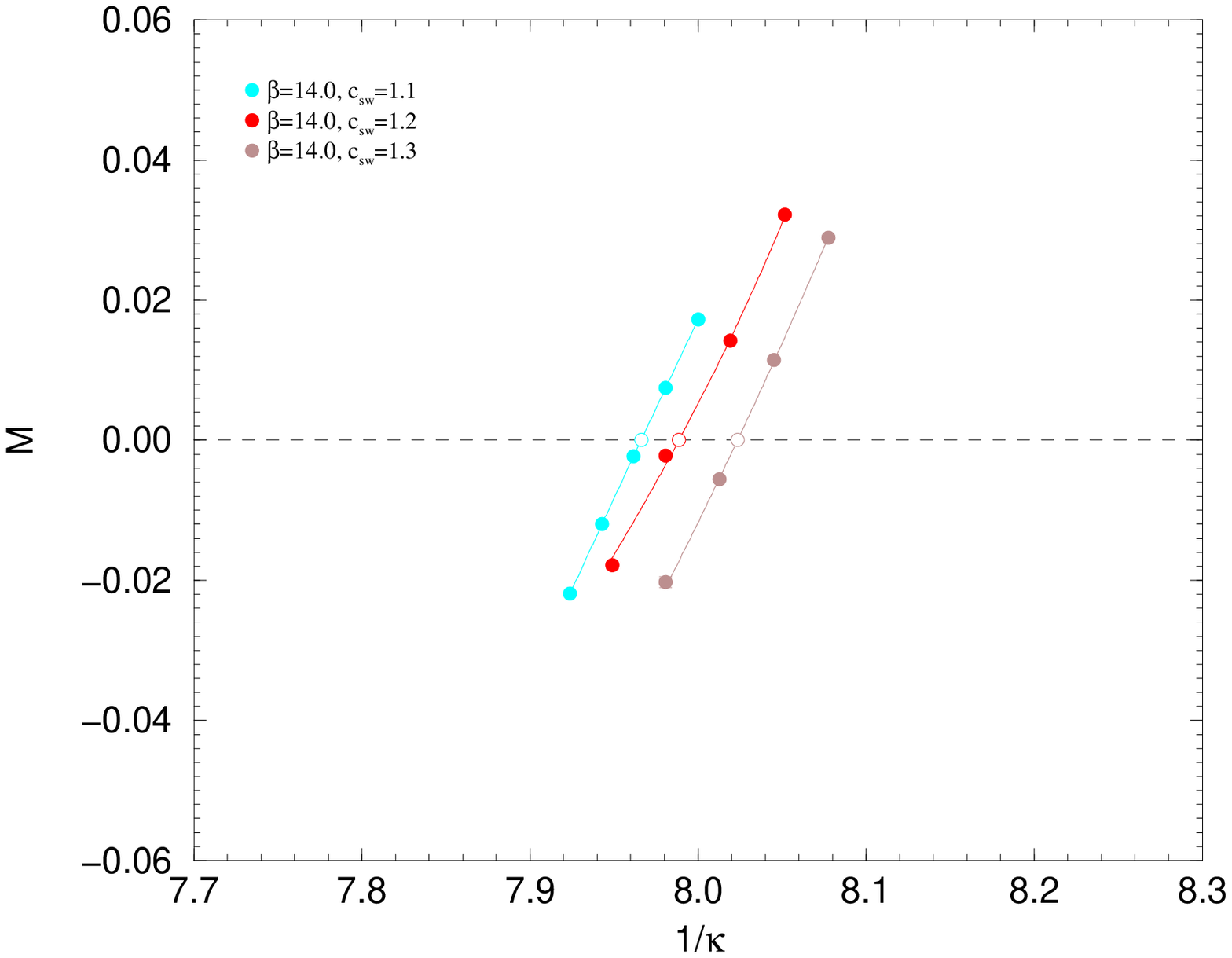}

\end{minipage}

\caption{$M$ against $1/\kappa$ for $\beta = 8.00$, $10.0$
         (upper left, right pictures respectively)
         and for $\beta = 14.0$,
         (lower picture),
         together with quadratic interpolations to $M = 0$
         (the open symbols).}

\label{ookap_M_b8p00-b14p0}

\end{figure}

\begin{figure}[p]

\begin{minipage}{0.475\textwidth}

   \epsfxsize=7.00cm \epsfbox{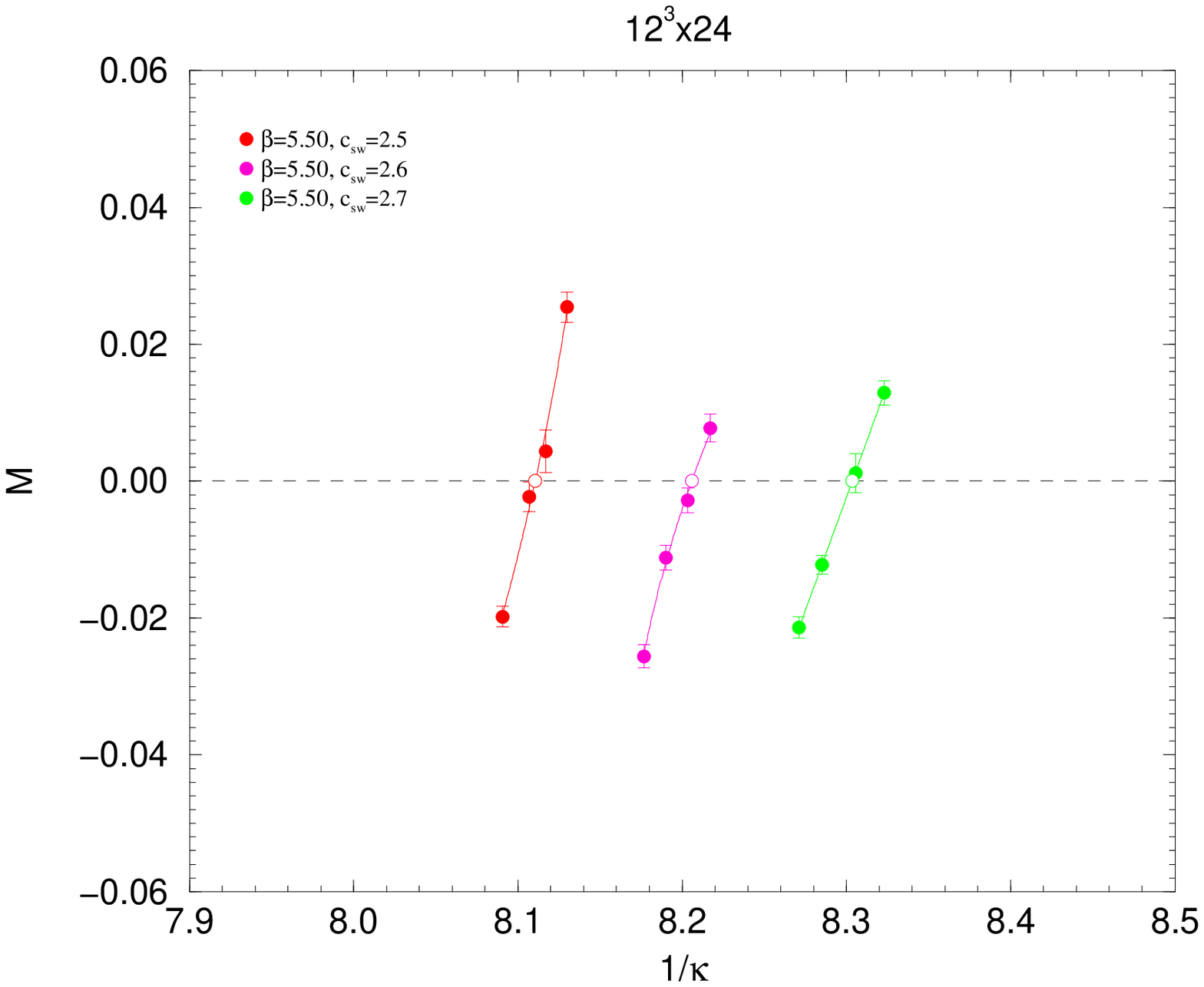}

\end{minipage} \hspace*{0.05\textwidth}
\begin{minipage}{0.475\textwidth}

   \epsfxsize=7.00cm \epsfbox{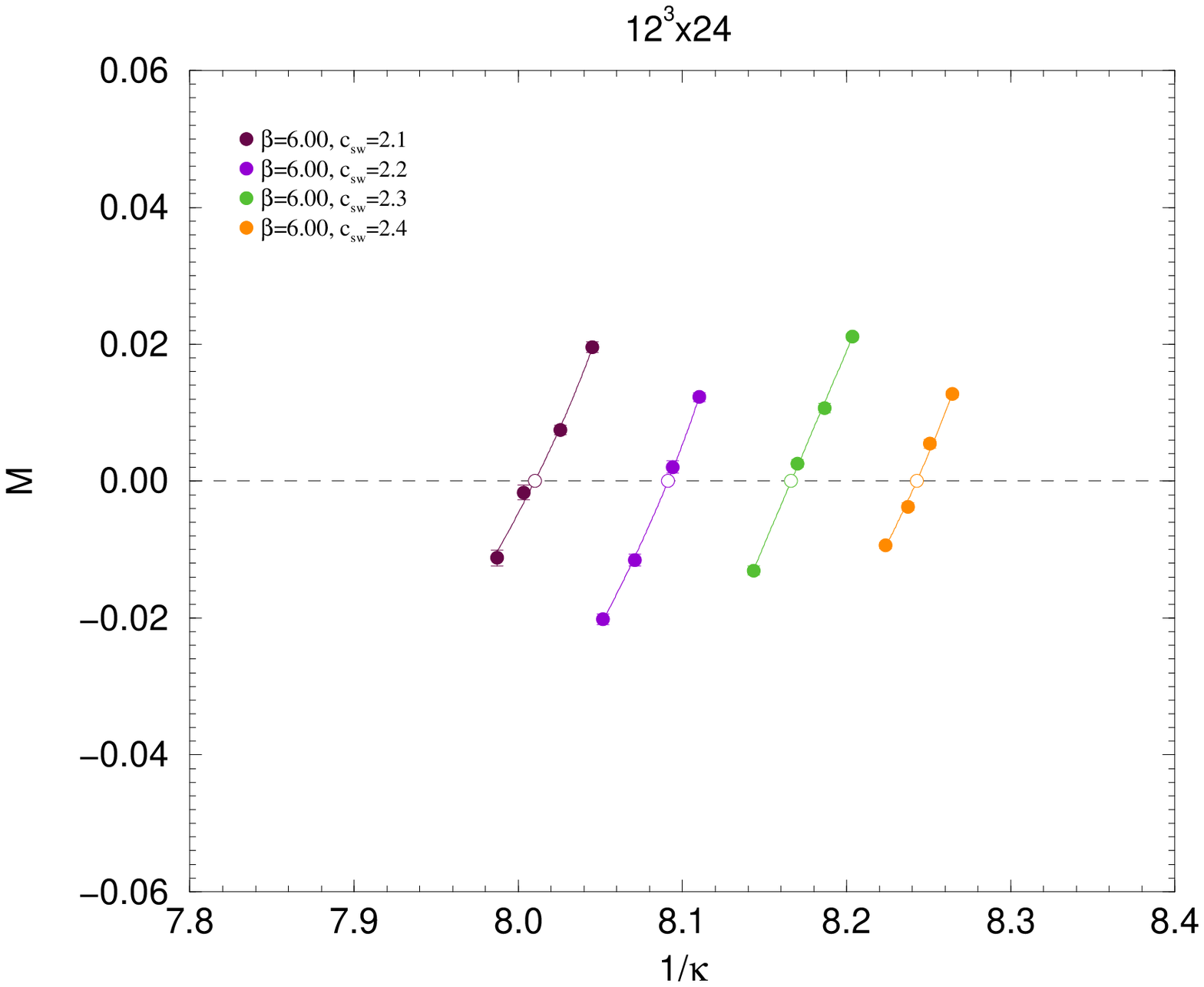}

\end{minipage}

\caption{$M$ against $1/\kappa$ for $\beta = 5.50$, $6.00$
         (left, right pictures respectively) on a $12^3\times 24$
         lattice together with quadratic interpolations to $M = 0$
         (the open symbols).}

\label{ookap-M_b5p50-b6p00_12x24}

\end{figure}

We first plot $M$ against $1/\kappa$ for the $8^3\times 16$ results in
Figs.~\ref{ookap_M_b5p10-b6p00}, \ref{ookap_M_b8p00-b14p0} and
for the $12^3\times 24$ results in Fig.~\ref{ookap-M_b5p50-b6p00_12x24}.

Note that to produce these graphs should not
require high statistics as it does not involve $\Delta M$. (Although
these are not the fundamental graphs they are also useful in helping
to determine the various $(c_{sw},\kappa)$ values for the runs.)

\begin{figure}[p]
\begin{minipage}{0.475\textwidth}

   \epsfxsize=7.00cm \epsfbox{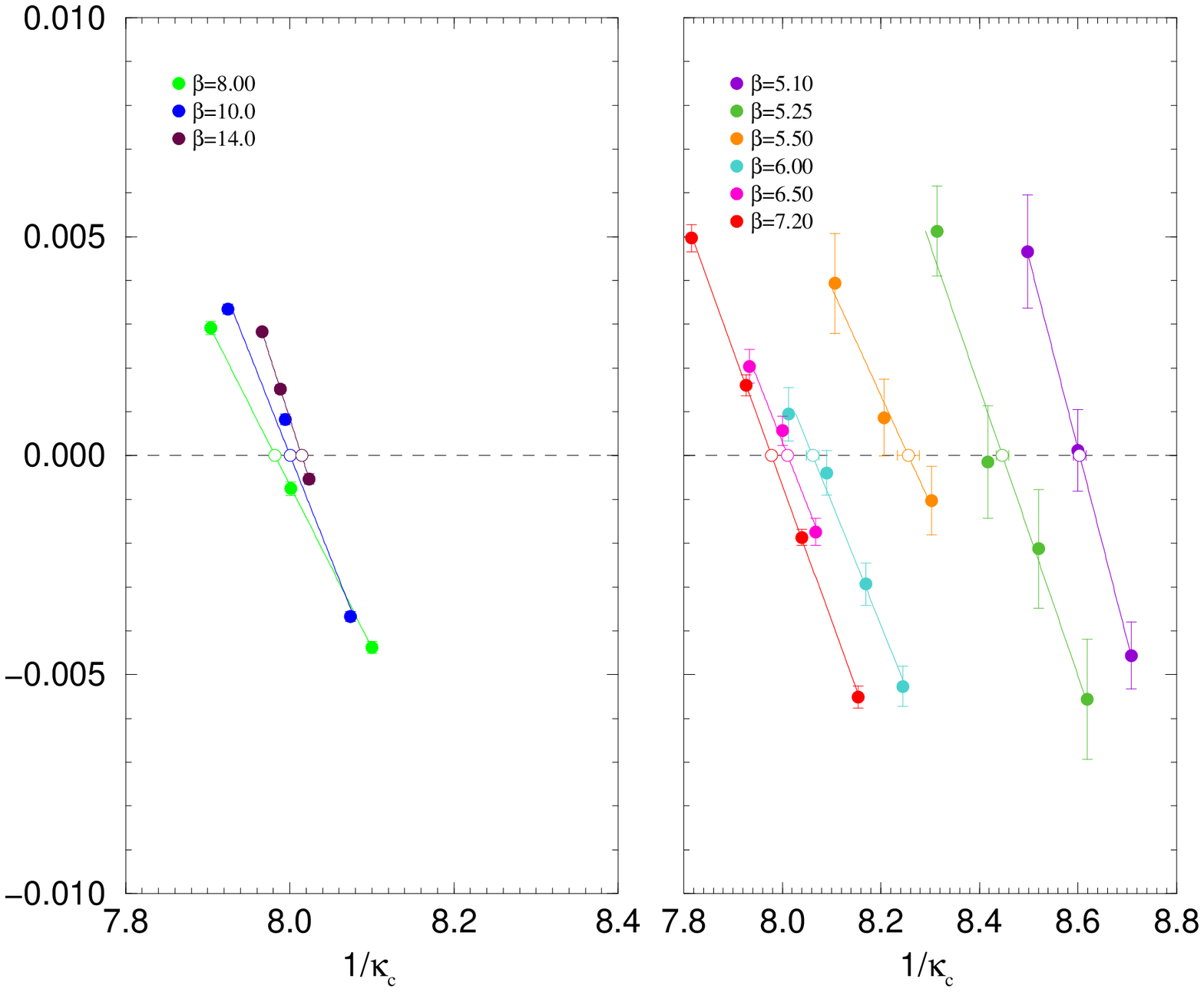}

\end{minipage} \hspace*{0.05\textwidth}
\begin{minipage}{0.475\textwidth}

   \epsfxsize=7.00cm \epsfbox{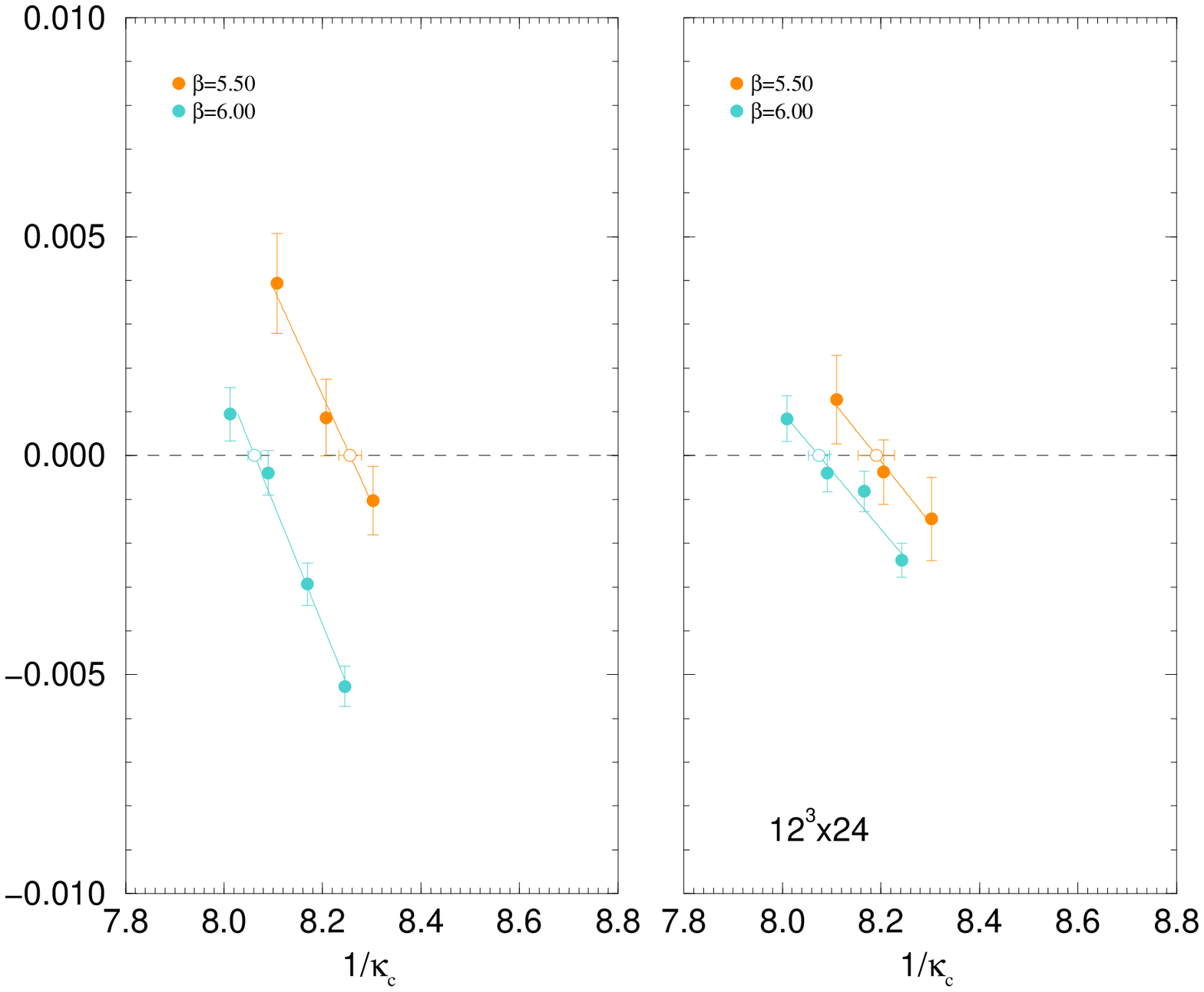}

\end{minipage}

\caption{Results of $\Delta M(\kappa_c(c_{sw}))$ versus
         $1/\kappa_c$ together with linear fits.
         The open circles give the optimal critical $\kappa_c$s,
         i.e.\ the $\kappa_c^*$s. The two left plots show the
         $8^3\times 16$ results while the two right plots
         compare the $\beta =5.50$, $6.00$ $8^3\times 16$ results
         with the $12^3\times 24$ results.}

\label{dM_pcac_ookapc_kapc}

\end{figure}

These $\Delta M(\kappa_c)$ are then plotted in 
Fig.~\ref{dM_pcac_ookapc_kapc}
again with a linear fit. Where $\Delta M$ vanishes
gives $\kappa_c^*$. For legibility the results have been
split into sub-graphs. We see that $\kappa_c^*$ is a non-monotonic
function of $\beta$.

We find results of
\begin{equation}
   \kappa_c^* = \left\{
      \begin{array}{l}
          \left. \begin{array}{lc}
                    0.116227(180) & \beta = 5.10 \\
                    0.118385(184) & \beta = 5.25 \\
                    0.121125(330) & \beta = 5.50 \\
                    0.124043(199) & \beta = 6.00 \\
                    0.124825(107) & \beta = 6.50 \\
                    0.125343(61)  & \beta = 7.20 \\
                    0.125281(38)  & \beta = 8.00 \\
                    0.124993(22)  & \beta = 10.0 \\
                    0.124773(26)  & \beta = 14.0 \\
                 \end{array}
          \right\} \, 8^3\times 16  \\
                                    \\
          \left. \begin{array}{lc}
                    0.122086(554) & \beta = 5.50 \\
                    0.123849(330) & \beta = 6.00 \\
                 \end{array}
          \right\} \, 12^3\times 24 \\
      \end{array}
               \right.
\label{kapcstar_vals}
\end{equation}

\begin{figure}[p]
   \hspace*{0.50in}
   \epsfxsize=10.50cm 
      \epsfbox{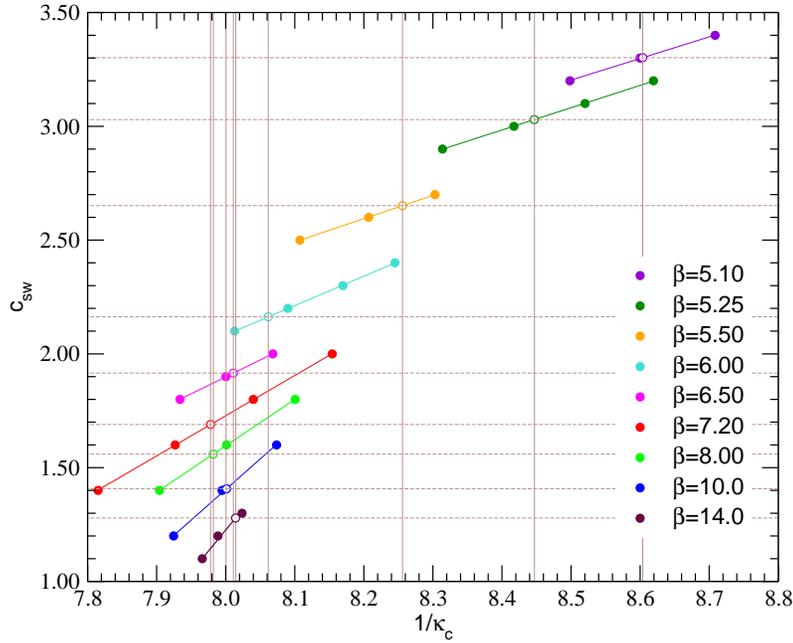}
   \caption{Results of $c_{sw}$ (filled circles) versus
            $1/\kappa_c$ together with linear fits.
            The optimal $c_{sw}$, $c_{sw}^*$, from 
            eq.~(\ref{cswstar_vals}) are shown as
            dashed horizontal lines. The open circles are
            the intersection of the linear fits with these horizontal
            lines and give an alternative determination of the optimal
            critical $\kappa_c$, $\kappa_c^*$, which are to be compared
            with the results of eq.~(\ref{kapcstar_vals}) shown as
            vertical lines.}
\label{ookapc_csw_b5p10-b14p0}
\end{figure}

As a consistency check the alternative plot of $c_{sw}$ against
$1/\kappa_c$ is shown in
Fig.~\ref{ookapc_csw_b5p10-b14p0}
where $c_{sw}$ is plotted against $1/\kappa_c(c_{sw})$,
again with a linear fit between the points. The optimal values of
$c_{sw}$, namely $c_{sw}^*$, taken from the previous fits as given
in eq.~(\ref{cswstar_vals}) are shown as dashed horizontal lines,
the intersection with the $1/\kappa_c$ curves then giving the optimal
critical values of $\kappa_c$, namely $\kappa_c^*$. These are denoted
in the figure as open points. As a comparison, the results from the
previous determination of $\kappa_c^*$, eq.~(\ref{kapcstar_vals}), are
also shown as vertical lines. We see good agreement between the different
determinations of $\kappa_c^*$, which indicates that the fit procedure
adopted here gives consistent results for both $c_{sw}^*$ and $\kappa_c^*$.
Finally note that plotting the $n_f=2$ flavour results would yield
a similar curve to Fig.~\ref{ookapc_csw_b5p10-b14p0}.

For future reference (in section~\ref{np_results}) as the fits in
Fig.~\ref{ookapc_csw_b5p10-b14p0} are all linear then we write
\begin{equation}
   {1 \over \kappa_c} 
      = {1 \over \kappa_c^*} + d ( c_{sw} - c_{sw}^* ) \,,
\label{d_def}
\end{equation}
with a measured coefficient $d(g_0)$,
\begin{equation}
   d = \left\{
      \begin{array}{l}
          \left. \begin{array}{lc}
                    1.0521(92)  & \beta = 5.10 \\
                    1.0208(54)  & \beta = 5.25 \\
                    0.9783(100) & \beta = 5.50 \\
                    0.7753(53)  & \beta = 6.00 \\
                    0.6722(51)  & \beta = 6.50 \\
                    0.5658(11)  & \beta = 7.20 \\
                    0.4907(10)  & \beta = 8.00 \\
                    0.3719(08)  & \beta = 10.0 \\
                    0.2704(23)  & \beta = 14.0 \\
                 \end{array}
          \right\} \, 8^3\times 16  \\
      \end{array}
       \right.
\label{grad_vals}
\end{equation}


\section{Finite size effects}
\label{finite_size_effects}


There are (small) ambiguities due to the finite volume used.
In an infinite volume we expect $O(a \Lambda_{\QCD})$ contributions
(in the chiral limit, otherwise there are also extra $O(am_q)$ terms)
due to the different boundary conditions or operators chosen.
In a finite volume there are additional $O(a/L)$ terms.
Thus might expect asymptotically, following \cite{aoki05a},
\begin{equation}
   c_{sw}^*(g_0, L/a)
      = c_{sw}^*(g_0,\infty) + c_L\,{a \over L} 
                              + c_\Lambda\, a\Lambda_{\QCD} + \ldots \,.
\label{finite_csw_kapc}
\end{equation}
The terms proportional to $a \Lambda_{\QCD}$ vanish as $a$ (or $g_0^2$)
$\to 0$ and represent the ambiguities in the different definitions of $M$.
For a physical quantity ${\cal Q}$, then 
\begin{eqnarray}
   {\cal Q} 
      &=& {\cal Q}(a) + q_L \,(c_{sw}^*(g_0, L/a) - c_{sw}^*(g_0, \infty))
                              \,a\Lambda_{\QCD} + O(a^2)
                                                              \nonumber  \\
      &=& {\cal Q}(a) + q_Lc_L\,{a \over L}\,a\Lambda_{\QCD}
                                                 + O(a^2) \,.
\label{finite_Q}
\end{eqnarray}
The correction term may be re-written as (where $L = aN_s$)
\begin{equation}
   q_Lc_L\,{a \over L}\,a\Lambda_{\QCD}
      = { q_Lc_L \over N_s}\, a\Lambda_{\QCD} \,.
\end{equation}
Potentially this might mean that ${\cal Q}$ is no longer $O(a)$ improved
for simulations where $c_{sw}^*$ has been determined on a fixed lattice
size, $N_s$. However it is likely that the unknown coefficients $q_L$
and $c_L$ are small and coupled with the $N_s$ factor in the denominator,
this is then expected to be a small effect.

To avoid this altogether we can either keep $L$ fixed in physical units
as $a \to 0$ (the `constant physics condition') so $O(a/L) \to 0$,
or alternatively simulate for several values of $N_s$ and extrapolate
to $N_s \to \infty$. The `Poor man's solution' is to evaluate at large
$\beta \to \infty$ (i.e.\ on a free configuration for $N_s=8$ here) and
subtract this result. Practically, following the same procedure as in
section~\ref{section_cswstar} we have found that for $c_{sw}$ this
$O(1/N_s)$ term (for $N_s = 8$) is negligible.

As noted previously we have also performed additional simulations on
larger lattices $12^3\times 24$ for $\beta = 6.00$, $5.50$ in order
to discuss finite lattice size corrections.
The results are plotted
in Figs.~\ref{M-dM_b5p50-b6p00_12x24}, \ref{ookap-M_b5p50-b6p00_12x24}
and compared with the $8^3\times 16$ results in
Figs.~\ref{csw_dM}, \ref{dM_pcac_ookapc_kapc}.
At tree level we have, \cite{luscher96c},
\begin{equation}
   \Delta M^{tree} = k\,(c_{sw}^{tree}  - 1){a \over L} + \ldots \,,
\end{equation}
which would indicate that for larger $N_s$ then $\Delta M$ becomes
smaller, with the consequent noise/signal ratio becoming worse.
Indeed this is seen in our results, with the $12^3\times 24$ data
being more bunched together in Fig.~\ref{M-dM_b5p50-b6p00_12x24}
than for the corresponding $8^3\times 16$ data in Fig.~\ref{M-dM_b5p10-b6p00}.
This may be mitigated somewhat by choosing a larger range of $c_{sw}$
due to the linear nature of the data as seen in 
Fig.~\ref{csw_dM} and eq.~(\ref{deltaM_csw_linear_fit}).
For $\beta = 6.00$ we have increased the number of $c_{sw}s$ used
in the analysis.

\begin{figure}[t]

\begin{minipage}{0.475\textwidth}

   \epsfxsize=7.00cm \epsfbox{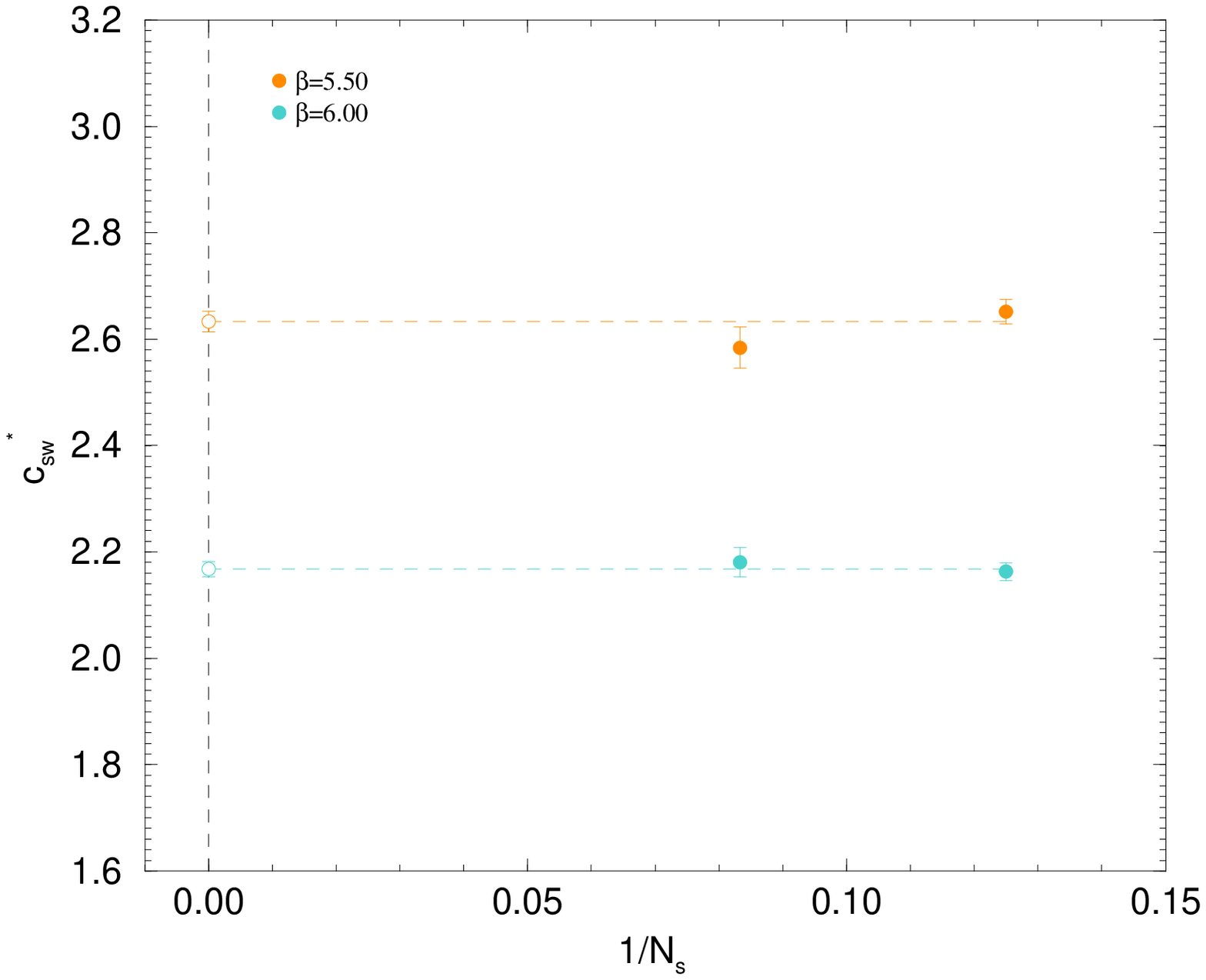}

\end{minipage} \hspace*{0.05\textwidth}
\begin{minipage}{0.475\textwidth}

   \epsfxsize=7.00cm \epsfbox{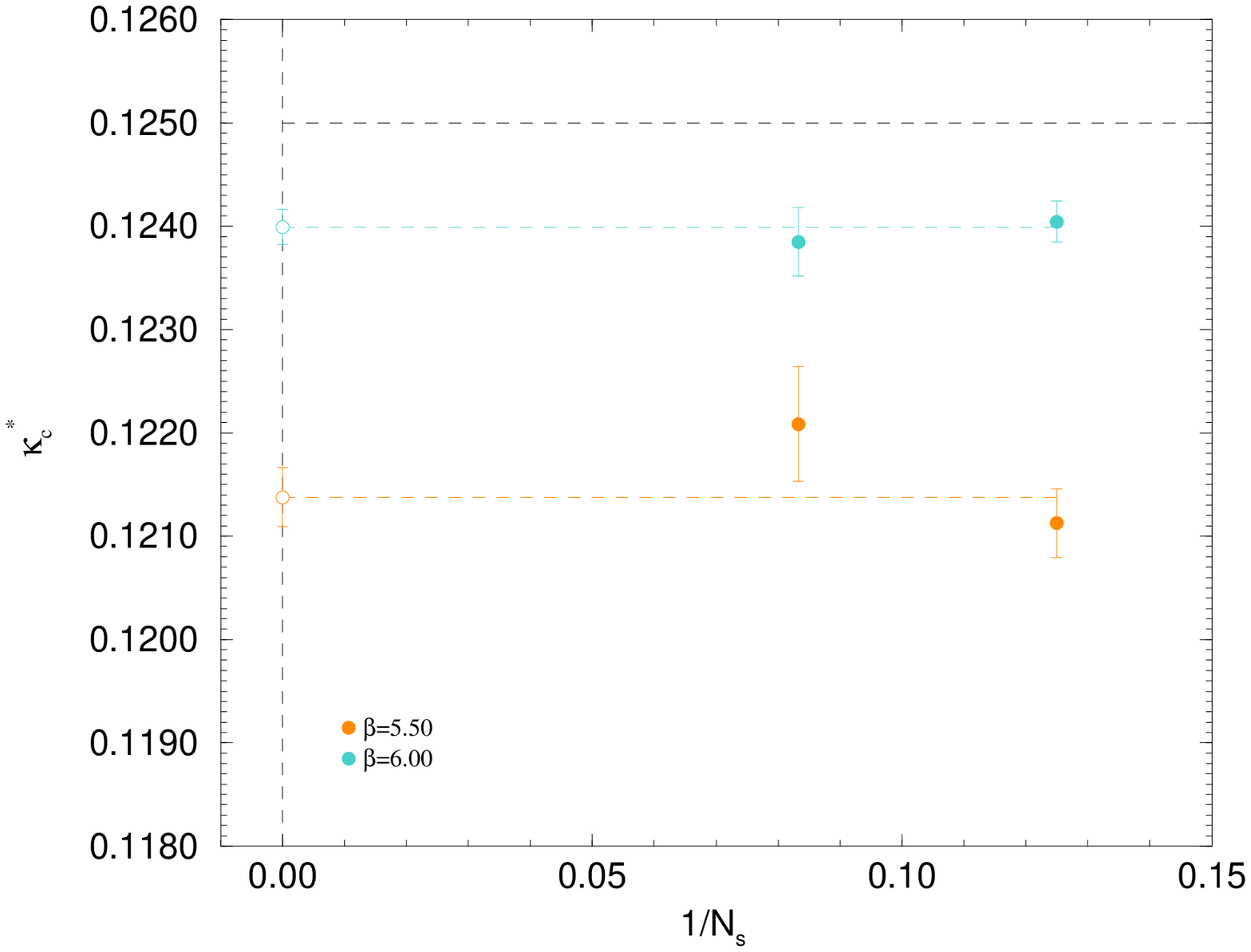}

\end{minipage}

\caption{$c_{sw}^*$ against $1/N_s$ (left picture) and
         $\kappa_c^*$ against $1/N_s$ (right picture)
         for $\beta = 5.50$, $6.00$, filled circles. Also shown
         are constant fits (dashed lines) together with the
         extrapolated values (open circles).}

\label{ooLs}

\end{figure}

In Fig.~\ref{ooLs} we plot $c_{sw}^*$ and $\kappa_c^*$ against
$1/N_s$. For both $\beta = 6.00$ and $5.50$ there seems to be small
finite size effects for $c_{sw}^*$. For $\kappa_c^*$ this is also the
case for $\beta = 6.00$, while for $\beta = 5.50$ the situation
is perhaps a little less clear-cut. However there is no systematic
trend in the data and a constant fit always lies within the error
bars of the data. So although we cannot come to a definite conclusion,
there do not seem to be large finite volume effects, i.e.\ $c_L$
appears to be small in eq.~(\ref{finite_csw_kapc}). So in
eq.~(\ref{finite_Q}) we only expect small violations of $O(a)$
improvement. We shall, in future, just consider the $8^3\times 16$ data.


\section{Results for $c_{sw}^*$ and $\kappa_c^*$}
\label{results}



\subsection{Perturbative results for $c_{sw}^*$ and $\kappa_c^*$}
\label{pert_results}


Before giving the non-perturbative results for $c_{sw}^*$ and $\kappa_c^*$
we first recapitulate the perturbative results. The lowest order perturbative
limit has been computed for both $c_{sw}^*$ and $\kappa_c^*$, \cite{perlt08a}.
For $c_{sw}^*$ we have
\begin{equation}
   c_{sw}^*(g_0) 
      = 1 + (0.196244 + 1.151888 \alpha - 4.2391365 \alpha^2)g_0^2 \,,
\end{equation}
where $\alpha$ is the stout smearing parameter, set equal to $0.1$ here.
This gives
\begin{equation}
   c_{sw}^*(g_0) = 1 + c_1 g_0^2 \,, \qquad c_1 = 0.269041 \,,
\label{csw_pert}
\end{equation}
i.e. the smearing parameter has increased the value
of $c_{sw}^*$ (for $\alpha =0$, we have $c_1 = 0.196244$).
For $\kappa_c(c_{sw}, g_0)$ we have
\begin{eqnarray}
   \kappa_c(c_{sw}, g_0)
              &=& {1 \over 8} 
                \left[ 1 +
                   \left( 0.0853699 - 0.961525\alpha + 3.55806\alpha^2
                \right.\right.
                                                                     \\
              & & \hspace*{0.50in} \left.\left.
                     - ( 0.025221 - 0.0787379\alpha) c_{sw} 
                      - 0.00984224c_{sw}^2  \right) g_0^2
                \right] \,,
                                                           \nonumber
\end{eqnarray}
giving for $\alpha = 0.1$
\begin{equation}
   \kappa_c(c_{sw}, g_0)
               = {1 \over 8} \left[ 1 + 
                 \left( 0.024798 - 0.0173472c_{sw} - 0.00984224c_{sw}^2
                 \right) g_0^2   \right] \,,
\label{kappac_pert_csw}
\end{equation}
and finally for $c_{sw} = c_{sw}^{tree} = 1$,
\begin{equation}
   \kappa_c^*(g_0) = {1 \over 8} \left[ 1 + k_1 g_0^2 \right] \,, \qquad
                      k_1 = - 0.002391 \,.
\label{kappac_pert}
\end{equation}
(Note that the result for $\kappa_c(c_{sw}, g_0)$ is more general than the
one given in \cite{perlt08a} when only the result for $c_{sw} = 1$ was given.)


\subsection{Non-perturbative results for $c_{sw}^*$ and $\kappa_c^*$}
\label{np_results}

\begin{figure}[t]
   \hspace{0.50in}
   \epsfxsize=12.00cm
      \epsfbox{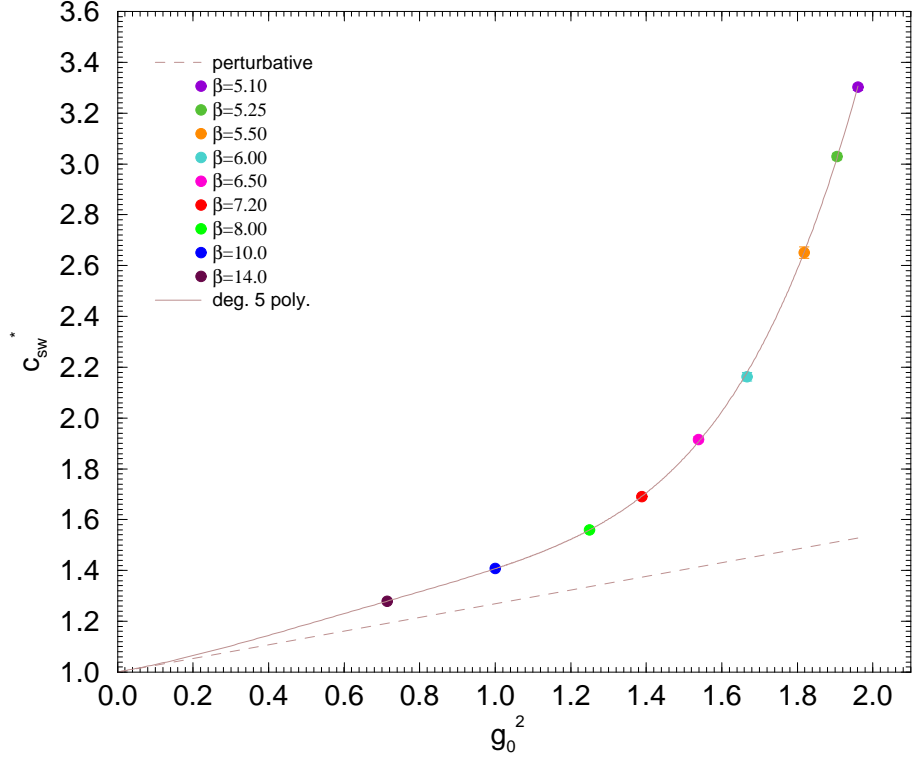}
   \caption{$c_{sw}^*$ against $g_0^2$ for various values of $\beta$
            (circles), together with a polynomial interpolation
            (line). Also shown is the perturbative result.}
\label{cswstar}
\end{figure}
\begin{figure}[t]
   \hspace{0.50in}
   \epsfxsize=12.00cm
      \epsfbox{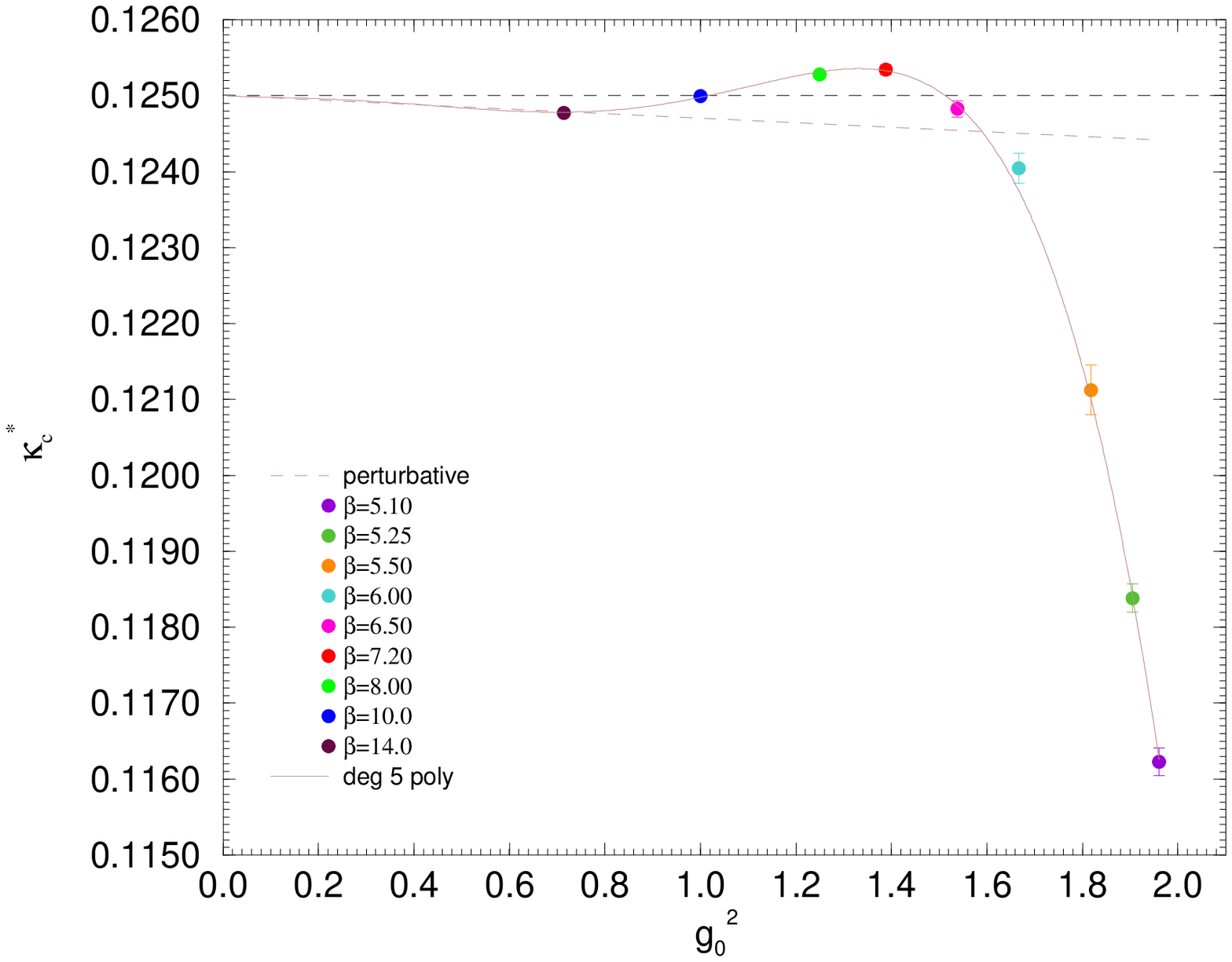}
   \caption{$\kappa_c^*$ against $g_0^2$ for various values of $\beta$
            (circles), together with a polynomial interpolation
            (line). Also shown is the perturbative result.}
\label{kapcstar}
\end{figure}

The results for $c_{sw}^*$ and $\kappa_c^*$ against $g_0^2$ are plotted
in Figs.~\ref{cswstar}, \ref{kapcstar} respectively in the range
$\beta \ge 5.10$. The lowest order perturbative limits are also
shown, eqs.~(\ref{csw_pert}) and (\ref{kappac_pert}).

An interpolation between the numerically determined
points is also shown. For both $c_{sw}^*$ and $\kappa_c^*$ a $5$th order
polynomial in $g_0^2$ proved sufficient. (These interpolation functions
are constrained to reproduce the perturbative results,
in the $\beta \to \infty$ limit and therefore, they have four free parameters.)
For $c_{sw}^*(g_0)$ we write
\begin{equation}
   c_{sw}^*(g_0) = 1 + c_1 g_0^2 + c_2 g_0^4 + c_3 g_0^6 + c_4 g_0^8
                     + c_5 g_0^{10} \,,
\label{cswstar_poly}
\end{equation}
and find
\begin{equation}
   \begin{array}{c|l}
      c_2 & +0.29910  \\
      c_3 & -0.11491  \\
      c_4 & -0.20003  \\
      c_5 & +0.15359  \\
   \end{array}
\end{equation}
while for $\kappa_c^*(g_0)$ we write
\begin{equation}
   \kappa_c^*(g_0) = {1 \over 8} \,
                     \left[ 1 + k_1 g_0^2 + k_2 g_0^4 + k_3 g_0^6 +
                                k_4 g_0^8 + k_5 g_0^{10} \right] \,,
\label{kapcstar_poly}
\end{equation}
and find
\begin{equation}
   \begin{array}{c|l}
      k_2 & +0.0122470 \\
      k_3 & -0.0525676 \\
      k_4 & +0.0668197 \\
      k_5 & -0.0242800 \\
   \end{array}
\end{equation}

These give for the specific $\beta$ values used here
\begin{equation}
   c_{sw}^* = \left\{ \begin{array}{lc}
                          3.306  & \beta = 5.10 \\
                          3.021  & \beta = 5.25 \\
                          2.653  & \beta = 5.50 \\
                          2.179  & \beta = 6.00 \\
                          1.907  & \beta = 6.50 \\
                          1.692  & \beta = 7.20 \\
                          1.560  & \beta = 8.00 \\
                          1.407  & \beta = 10.0 \\
                          1.279  & \beta = 14.0 \\
                      \end{array}
              \right. \qquad
   \kappa_c^* = \left\{ \begin{array}{lc}
                          0.116262 & \beta = 5.10 \\
                          0.118424 & \beta = 5.25 \\
                          0.120996 & \beta = 5.50 \\
                          0.123751 & \beta = 6.00 \\
                          0.124870 & \beta = 6.50 \\
                          0.125328 & \beta = 7.20 \\
                          0.125314 & \beta = 8.00 \\
                          0.124979 & \beta = 10.0 \\
                          0.124783 & \beta = 14.0 \\
                      \end{array}
              \right. \qquad
\label{cswstar_pade}
\end{equation}
which are to be compared with the numerically determined values. 
The errors for $c_{sw}^*$ from the fit are estimated to be
about $0.4\%$ while for $\kappa_c^*$ we have $0.02\%$
at $\beta = 14.0$ rising to $0.15\%$ at $\beta = 5.10$.

These smooth fits between the points give estimates for $c_{sw}^*$
(and $\kappa_c^*$) which could be used in the action for future generation
of configurations.

For $c_{sw}^*$ the polynomial only tracks the perturbative solution
for small values of $g_0^2$. This is perhaps not surprising as
the tadpole improved, $TI$, estimate is $c_{sw}^{\ti} = u_0^{(S)}/u_0^4$,
\cite{perlt08a}, which is to be compared with the unsmeared case
of $c_{sw}^{\ti} = 1/u_0^3$ where $u_0$ is the average plaquette value
and $u_0^{(S)}$ is the smeared value. As smearing increases
the plaquette value this indicates that $c_{sw}^*$ can be large.
For $\kappa_c^*$ on the other hand as $\kappa_c^{\ti} = 1/(8u_0^{(S)})$
we expect that it is $\sim 1/8$. This is true for reasonably fine lattices,
however $\kappa_c^*$ does begin to decrease for larger values of $g_0^2$.
For $n_f=2$ the same phenomenon occurs: for larger $g_0^2$,
$\kappa_c^*$ begins to decrease (after initially increasing).

As a further consistency check on the results, we can investigate the
gradient $\partial (1/\kappa_c) / \partial c_{sw}|_{c_{sw}^*}$.
From eq.~(\ref{d_def}) we have
\begin{equation}
   \left. {\partial (1/\kappa_c) \over \partial c_{sw}} 
            \right|_{c_{sw}^*} = d \,,
\end{equation}
as the fits in Fig.~\ref{ookapc_csw_b5p10-b14p0} are linear,
where $d$ is given in eq.~(\ref{grad_vals}).
Perturbatively we have from eq.~(\ref{kappac_pert_csw}),
\begin{equation}
   {\partial (1/\kappa_c) \over \partial c_{sw}} 
        = 8 \left[ 0.037032 + 0.019684(c_{sw}-1) \right] g_0^2 \,.
\end{equation}
As $g_0$ increases $c_{sw}$ increases, so not only do more terms
in this expansion become important,
but the coefficient of the leading term increases as well.
For $c_{sw} = c_{sw}^{tree} = 1$ we have the leading order perturbative result,
\begin{equation}
   \left. {\partial (1/\kappa_c) \over \partial c_{sw}} \right|_{c_{sw}^*}
        = d_1 g_0^2 \,, \qquad d_1 = 0.296253 \,.
\end{equation}

\begin{figure}[t]
   \hspace{0.50in}
   \epsfxsize=12.00cm
      \epsfbox{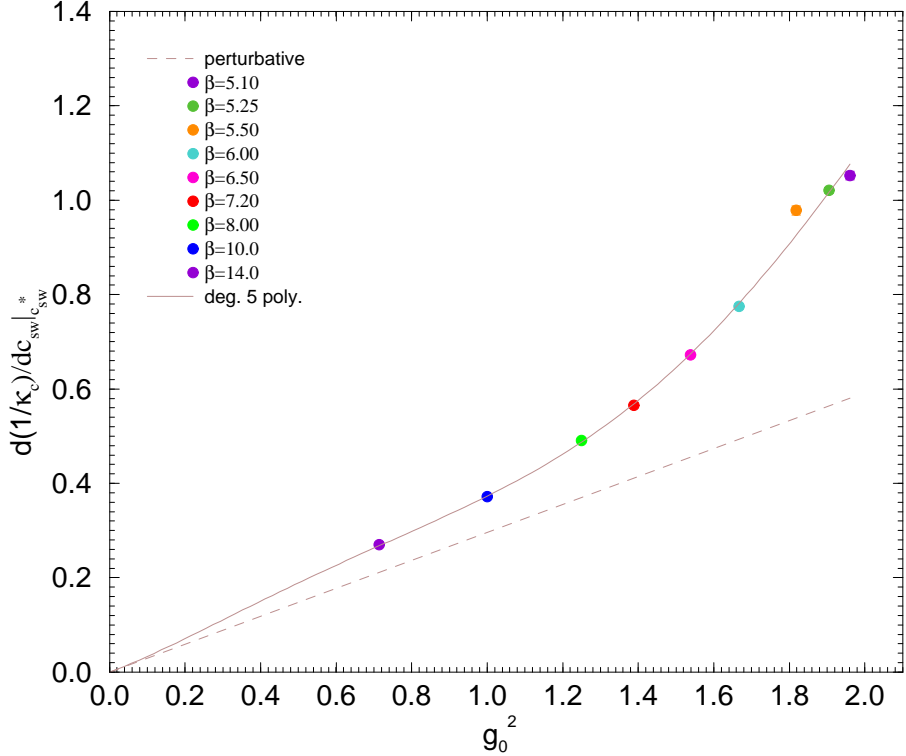}
   \caption{$\partial (1/\kappa_c) / \partial c_{sw}|_{c_{sw}^*}$
            against $g_0^2$ for various values of $\beta$
            (circles), together with a polynomial interpolation
            (line). Also shown is the perturbative result.}
\label{g2_dookapcodcsw}
\end{figure}

In Fig.~\ref{g2_dookapcodcsw} we plot
$\partial (1/\kappa_c) / \partial c_{sw}|_{c_{sw}^*}$ against
$g_0^2$, together with a $5$th order polynomial in $g_0^2$,
\begin{equation}
   \left. {\partial (1/\kappa_c) \over \partial c_{sw}} \right|_{c_{sw}^*}
     = d_1 g_0^2 + d_2 g_0^4 + d_3 g_0^6 + d_4 g_0^8 + d_5 g_0^{10} \,,
\end{equation}
and find
\begin{equation}
   \begin{array}{c|l}
      d_2 & +0.4180  \\
      d_3 & -0.7232  \\ 
      d_4 & +0.4739  \\
      d_5 & -0.0919  \\
   \end{array}
\end{equation}
The results follow a smooth curve.


\section{Conclusions and Discussion}
\label{conclusions}


Non-perturbative $O(a)$ improvement is a viable procedure for (stout)
smeared actions with typical clover results being obtained.
(Other recent results for $3$ flavours are given in
\cite{yamada04a,aoki05a,edwards08a}.) Using the Schr\"odinger
functional method we have determined the optimal clover coefficent,
$c_{sw}^*$ necessary to achieve $O(a)$ improvement and also the
optimal critical hopping parameter, $\kappa_c^*$, eqs.~(\ref{cswstar_poly}),
(\ref{kapcstar_poly}) over a wide range of coupling constant.

As $a$ increases we need a significant $c_{sw} \gg c_{sw}^{tree} \equiv 1$
for $O(a)$ improvement. We are now seeking a region where
$a \sim 0.05 \, - \, 0.1\, \mbox{fm}$. Improvement, which is presumably
represented by an asymptotic series, brings an advantage for smaller $a$
say $a \le 0.1\,\mbox{fm}$. The two extremes for $a$ are simulations at
small $a$ with `large' $m_{ps}$ when there is no continuum
extrapolation but a chiral extrapolation, or alternatively
simulations at `coarse' $a$ with $m_{ps} \sim m_{\pi}$
when there is no chiral extrapolation but a continuum extrapolation.
Of course the Schr\"odinger functional does not tell us $a$;
for this conventional HMC simulations are required. Some preliminary results
indicate that around $\beta \sim 5.50$ we have $a \sim 0.08\,\mbox{fm}$.


\section*{Acknowledgements}


The numerical calculations have been performed on the
BlueGeneLs at EPCC (Edinburgh, UK), NIC (J\"ulich, Germany),
the QCDOC (Edinburgh, UK) and the SGI ICE at HLRN
(Berlin-Hannover, Germany). We thank all institutions for their support.
The Chroma software library was used, \cite{edwards04a},
and we are grateful to R.~G. Edwards and B. Jo{\'o} for their help and advice.
The BlueGene and QCDOC codes were optimised using BAGEL, \cite{boyle05a}.
This work has been supported in part by
the EU Integrated Infrastructure Initiative Hadron Physics (I3HP) under
contract RII3-CT-2004-506078, by the DFG under contracts
FOR 465 (Forschergruppe Gitter-Hadronen-Ph\"anomenologie) and
SFB/TR 55 (Hadron Physics from Lattice QCD) and the HPC-EUROPA++ project
(project number 211437), funded by the European Community's Research
Infrastructure Action within the FP7 ``Coordination and support action''
Programme. JMZ acknowledges support from STFC Grant PP/F009658/1.


\clearpage

\appendix

\section*{Appendix}


\section{$M$ and $\Delta M$ results}
\label{appendix_raw_results}

We collect here in Tables~\ref{table_run_b5p10}, \ref{table_run_b5p25}, 
\ref{table_run_b5p50}, \ref{table_run_b6p00}, \ref{table_run_b6p50},
\ref{table_run_b7p20}, \ref{table_run_b8p00}, \ref{table_run_b10p0}
and \ref{table_run_b14p0} the numerical values of $M$, $\Delta M$
as defined in eq.~(\ref{M+dM_def}) for the
$N_s^3\times 2N_s = 8^3\times 16$ lattices, while in
tables~\ref{table_run_5p50_12x24} and \ref{table_run_6p00_12x24}
the results for the $12^3 \times 24$ lattices are given.

The data sets are of size $O(3000)$ trajectories for the
$8^3\times 16$ lattices and $O(2000)$ trajectories
for the $12^3\times 24$ lattices. An initial thermalisation phase
was typically of order $300$ trajectories. The trajectory length
$\tau_{chroma}$ was always $1$, while the number of steps in the
trajectory, $n_{\tau_{chroma}}$, varied for the $8^3\times 16$ lattices
from $10$ for $\beta \ge 6.50$ to $12$, $12$, $15$, $18$ for
$\beta = 6.00$, $5.50$, $5.25$, $5.10$ respectively. This maintained
an acceptance rate of $ > 80\%$. (This decreased very slightly
for the larger $\beta$-values.) For the $12^3\times 24$ lattices
$n_{\tau_{chroma}} = 18$, $22$ for $\beta = 6.00$, $5.50$ was used to
give this acceptance. 

The jackknife errors for the ratios are given uniformly
to two significant figures, with the overriding requirement
that the result must also have a minimum of four significant figures.
To reduce possible autocorrelations in the data every second trajectory
was used with a jackknife block size of $10$.

\vspace*{0.50in}


\begin{table}[h]
   \begin{center}
      \begin{tabular}{||l|l|l||l|l||}
         \hline
         \hline
\multicolumn{1}{||c}{$\beta$}                               &
\multicolumn{1}{|c}{$c_{sw}$}                               &
\multicolumn{1}{|c||}{$\kappa$}                             &
\multicolumn{1}{c}{$M$}                                     &
\multicolumn{1}{|c||}{$\Delta M$}                           \\
         \hline
 5.10 & 3.20 & 0.11760 & \phantom{-}0.007049(2313) &\phantom{-}0.005762(1923) \\
 5.10 & 3.20 & 0.11780 & -0.01315(205)   & \phantom{-}0.001565(1545) \\
 5.10 & 3.20 & 0.11800 & -0.02324(231)   & \phantom{-}0.003037(1212) \\
 5.10 & 3.20 & 0.11820 & -0.04187(212)   & -0.001353(1782) \\
         \hline
 5.10 & 3.30 & 0.11610 & \phantom{-}0.01941(227)  &\phantom{-}0.004648(1570) \\
 5.10 & 3.30 & 0.11620 & \phantom{-}0.001408(2298) & -0.003942(1737) \\
 5.10 & 3.30 & 0.11640 & -0.005654(2058) & \phantom{-}0.001279(1438) \\
 5.10 & 3.30 & 0.11660 & -0.02596(166)   & -0.002347(1310) \\
 5.10 & 3.30 & 0.11690 & -0.04356(181)   & -0.004137(1550) \\
         \hline
 5.10 & 3.40 & 0.11470 & \phantom{-}0.01098(191)   & -0.003299(1305) \\
 5.10 & 3.40 & 0.11490 & -0.004606(1516) & -0.004438(1044) \\
 5.10 & 3.40 & 0.11510 & -0.01742(160)   & -0.006135(1442) \\
 5.10 & 3.40 & 0.11530 & -0.02432(125)   & -0.005855(748)  \\
 5.10 & 3.40 & 0.11550 & -0.03424(165)   & -0.004780(1086) \\
         \hline
         \hline
 \end{tabular}
   \end{center}
\caption{$8^3\times 16$ results for $M$ and $\Delta M$ for $\beta = 5.10$.}
\label{table_run_b5p10}
\end{table}

\clearpage


\begin{table}[t]
   \begin{center}
      \begin{tabular}{||l|l|l||l|l||}
         \hline
         \hline
\multicolumn{1}{||c}{$\beta$}                               &
\multicolumn{1}{|c}{$c_{sw}$}                               &
\multicolumn{1}{|c||}{$\kappa$}                             &
\multicolumn{1}{c}{$M$}                                     &
\multicolumn{1}{|c||}{$\Delta M$}                           \\
         \hline
 5.25 & 2.90 & 0.12000 & \phantom{-}0.02772(322)   &\phantom{-}0.007506(2096) \\
 5.25 & 2.90 & 0.12015 & \phantom{-}0.01468(226)  & \phantom{-}0.008527(1944) \\
 5.25 & 2.90 & 0.12025 & \phantom{-}0.005850(2967) &\phantom{-}0.003412(1559) \\
 5.25 & 2.90 & 0.12050 & -0.02097(186)   & \phantom{-}0.004020(1167) \\
 5.25 & 2.90 & 0.12100 & -0.04947(241)   & \phantom{-}0.0008952(14184)        \\
         \hline
 5.25 & 3.00 & 0.11860 & \phantom{-}0.02041(279)   & -0.0007319(14965)        \\
 5.25 & 3.00 & 0.11875 &\phantom{-}0.0008556(19694)&\phantom{-}0.001173(1115) \\
 5.25 & 3.00 & 0.11890 & -0.006210(2160) & -0.001295(1424) \\
 5.25 & 3.00 & 0.11905 & -0.01727(244)   & -0.006479(2808) \\
 5.25 & 3.00 & 0.11920 & -0.03280(169)   & -0.002655(1102) \\
         \hline
 5.25 & 3.10 & 0.11700 & \phantom{-}0.01973(153)   & -0.001642(991)  \\
 5.25 & 3.10 & 0.11720 & \phantom{-}0.01021(171)   & -0.002551(1054) \\
 5.25 & 3.10 & 0.11740 & -0.002194(1506) & -0.001440(922)  \\
 5.25 & 3.10 & 0.11760 & -0.01303(133)   & -0.002512(1050) \\
 5.25 & 3.10 & 0.11780 & -0.02344(186)   & -0.0008732(12753) \\
         \hline
 5.25 & 3.20 & 0.11580 & \phantom{-}0.01019(131)   & -0.005591(815)  \\
 5.25 & 3.20 & 0.11600 & \phantom{-}0.0001673(11516)& -0.005485(875)  \\
 5.25 & 3.20 & 0.11620 & -0.008058(1185) & -0.005259(1342) \\
 5.25 & 3.20 & 0.11640 & -0.01905(114)   & -0.003621(1214) \\
         \hline
         \hline
 \end{tabular}
   \end{center}
\caption{$8^3\times 16$ results for $M$ and $\Delta M$ for $\beta = 5.25$.}
\label{table_run_b5p25}
\end{table}


\begin{table}[t]
   \begin{center}
      \begin{tabular}{||l|l|l||l|l||}
         \hline
         \hline
\multicolumn{1}{||c}{$\beta$}                               &
\multicolumn{1}{|c}{$c_{sw}$}                               &
\multicolumn{1}{|c||}{$\kappa$}                             &
\multicolumn{1}{c}{$M$}                                     &
\multicolumn{1}{|c||}{$\Delta M$}                           \\
         \hline
 5.50 & 2.50 & 0.12300 & \phantom{-}0.02608(208)   &\phantom{-}0.005685(1134) \\
 5.50 & 2.50 & 0.12320 & \phantom{-}0.01112(215)  &\phantom{-}0.003630(1484) \\
 5.50 & 2.50 & 0.12335 & -0.001449(2014) & \phantom{-}0.004018(1567) \\
 5.50 & 2.50 & 0.12360 & -0.01565(210)   & \phantom{-}0.007337(1321) \\
         \hline
 5.50 & 2.60 & 0.12170 &\phantom{-}0.006007(1703)&\phantom{-}0.0009700(16220) \\
 5.50 & 2.60 & 0.12190 & -0.0001614(18320) &\phantom{-}0.001739(1046) \\
 5.50 & 2.60 & 0.12210 & -0.01343(170)   & 0.0001628(11051)          \\
 5.50 & 2.60 & 0.12230 & -0.01959(223)   & \phantom{-}0.003397(1508) \\
         \hline
 5.50 & 2.70 & 0.12015 & \phantom{-}0.01584(149)   & -0.002008(1139) \\
 5.50 & 2.70 & 0.12040 & \phantom{-}0.002419(1200) & -0.001062(798)  \\
 5.50 & 2.70 & 0.12070 & -0.01264(125)   & -0.001321(1175) \\
 5.50 & 2.70 & 0.12090 & -0.01831(150)   & -0.001626(915)  \\
         \hline
         \hline
 \end{tabular}
   \end{center}
\caption{$8^3\times 16$ results for $M$ and $\Delta M$ for $\beta = 5.50$.}
\label{table_run_b5p50}
\end{table}

\clearpage


\begin{table}[t]
   \begin{center}
      \begin{tabular}{||l|l|l||l|l||}
         \hline
         \hline
\multicolumn{1}{||c}{$\beta$}                               &
\multicolumn{1}{|c}{$c_{sw}$}                               &
\multicolumn{1}{|c||}{$\kappa$}                             &
\multicolumn{1}{c}{$M$}                                     &
\multicolumn{1}{|c||}{$\Delta M$}                           \\
         \hline
 6.00 & 2.10 & 0.12430 & \phantom{-}0.01841(99)    &\phantom{-}0.001623(800)  \\
 6.00 & 2.10 & 0.12460 & \phantom{-}0.006443(1084) &\phantom{-}0.001332(753)  \\
 6.00 & 2.10 & 0.12495 & -0.004446(970)  & \phantom{-}0.0006452(7878) \\
 6.00 & 2.10 & 0.12520 & -0.01316(107)   &\phantom{-}0.003539(970)  \\
         \hline
 6.00 & 2.20 & 0.12330 & \phantom{-}0.01135(86)    & -0.0007576(5905) \\
 6.00 & 2.20 & 0.12355 & \phantom{-}0.002234(706)  & -0.0001747(6084) \\
 6.00 & 2.20 & 0.12390 & -0.01050(79)    & -0.0008061(7138) \\
 6.00 & 2.20 & 0.12420 & -0.02108(79)    & -0.0008650(6771) \\
         \hline
 6.00 & 2.30 & 0.12190 & \phantom{-}0.01996(58)    & -0.002989(439)  \\
 6.00 & 2.30 & 0.12215 & \phantom{-}0.009817(838)  & -0.002765(574)  \\
 6.00 & 2.30 & 0.12240 & \phantom{-}0.0001335(7744) & -0.003061(672)  \\
 6.00 & 2.30 & 0.12280 & -0.01430(67)    & -0.003268(549)  \\
         \hline
 6.00 & 2.40 & 0.12100 & \phantom{-}0.01228(69)    & -0.004705(456)  \\
 6.00 & 2.40 & 0.12120 & \phantom{-}0.003415(610)  & -0.005526(586)  \\
 6.00 & 2.40 & 0.12140 & -0.004357(723)  & -0.004751(540)  \\
 6.00 & 2.40 & 0.12160 & -0.01066(73)    & -0.004149(657)  \\
         \hline
         \hline
 \end{tabular}
   \end{center}
\caption{$8^3\times 16$ results for $M$ and $\Delta M$ for $\beta = 6.00$.}
\label{table_run_b6p00}
\end{table}


\begin{table}[t]
   \begin{center}
      \begin{tabular}{||l|l|l||l|l||}
         \hline
         \hline
\multicolumn{1}{||c}{$\beta$}                               &
\multicolumn{1}{|c}{$c_{sw}$}                               &
\multicolumn{1}{|c||}{$\kappa$}                             &
\multicolumn{1}{c}{$M$}                                     &
\multicolumn{1}{|c||}{$\Delta M$}                           \\
         \hline
 6.50 & 1.80 & 0.12550 & \phantom{-}0.01994(59)    &\phantom{-}0.001612(472)  \\
 6.50 & 1.80 & 0.12575 & \phantom{-}0.01067(59)    &\phantom{-}0.001914(457)  \\
 6.50 & 1.80 & 0.12600 & \phantom{-}0.001513(513)  &\phantom{-}0.001973(466)  \\
 6.50 & 1.80 & 0.12650 & -0.01600(55)    & \phantom{-}0.002172(496)  \\
         \hline
 6.50 & 1.90 & 0.12440 & \phantom{-}0.02139(60)   &\phantom{-}0.0004039(4011) \\
 6.50 & 1.90 & 0.12470 & \phantom{-}0.01068(56)    &\phantom{-}0.001113(435)  \\
 6.50 & 1.90 & 0.12495 & \phantom{-}0.001754(539) &\phantom{-}0.0003388(5215) \\
 6.50 & 1.90 & 0.12520 & -0.007849(601)  & -0.00003026(52328)  \\
         \hline
 6.50 & 2.00 & 0.12360 & \phantom{-}0.01255(49)    & -0.002074(450)  \\
 6.50 & 2.00 & 0.12390 & \phantom{-}0.001931(525)  & -0.001253(358)  \\
 6.50 & 2.00 & 0.12410 & -0.006006(505)  & -0.002711(510)  \\
 6.50 & 2.00 & 0.12440 & -0.01635(49)    & -0.001294(453)  \\
         \hline
         \hline
 \end{tabular}
   \end{center}
\caption{$8^3\times 16$ results for $M$ and $\Delta M$ for $\beta = 6.50$.}
\label{table_run_b6p50}
\end{table}

\clearpage


\begin{table}[t]
   \begin{center}
      \begin{tabular}{||l|l|l||l|l||}
         \hline
         \hline
\multicolumn{1}{||c}{$\beta$}                               &
\multicolumn{1}{|c}{$c_{sw}$}                               &
\multicolumn{1}{|c||}{$\kappa$}                             &
\multicolumn{1}{c}{$M$}                                     &
\multicolumn{1}{|c||}{$\Delta M$}                           \\
         \hline
 7.20 & 1.40 & 0.12720 & \phantom{-}0.02534(46)    &\phantom{-}0.006503(387)  \\
 7.20 & 1.40 & 0.12797 & -0.0007597(4109) & \phantom{-}0.005029(430)  \\
 7.20 & 1.40 & 0.12850 & -0.01713(48)    &  \phantom{-}0.005118(372)  \\
 7.20 & 1.40 & 0.12920 & -0.03970(53)    &  \phantom{-}0.007053(563)  \\
         \hline
 7.20 & 1.60 & 0.12500 & \phantom{-}0.03839(43)    &\phantom{-}0.003053(391)  \\
 7.20 & 1.60 & 0.12570 & \phantom{-}0.01500(38)    &\phantom{-}0.001534(389)  \\
 7.20 & 1.60 & 0.12615 & \phantom{-}0.0003391(4484)&\phantom{-}0.001883(311)  \\
 7.20 & 1.60 & 0.12660 & -0.01525(36)    &  \phantom{-}0.001353(543)  \\
 7.20 & 1.60 & 0.12720 & -0.03608(38)    &  \phantom{-}0.001644(307)  \\
         \hline
 7.20 & 1.80 & 0.12270 & \phantom{-}0.05607(29)    & -0.001786(239)  \\
 7.20 & 1.80 & 0.12380 & \phantom{-}0.01959(32)    & -0.001553(260)  \\
 7.20 & 1.80 & 0.12438 & -0.00008070(34186)        & -0.002103(288)  \\
 7.20 & 1.80 & 0.12500 & -0.02136(34)    & -0.001939(300)  \\
 7.20 & 1.80 & 0.12590 & -0.05319(35)    & -0.001455(388)  \\
         \hline
 7.20 & 2.00 & 0.12150 & \phantom{-}0.03819(31)    & -0.005604(315)  \\
 7.20 & 2.00 & 0.12210 & \phantom{-}0.01736(38)    & -0.005245(340)  \\
 7.20 & 2.00 & 0.12264 & -0.0002027(3196)          & -0.005262(470)  \\
 7.20 & 2.00 & 0.12290 & -0.008518(356)            & -0.005990(375)  \\
 7.20 & 2.00 & 0.12360 & -0.03421(34)              & -0.006188(311)  \\

         \hline
         \hline
 \end{tabular}
   \end{center}
\caption{$8^3\times 16$ results for $M$ and $\Delta M$ for $\beta = 7.20$.}
\label{table_run_b7p20}
\end{table}


\begin{table}[t]
   \begin{center}
      \begin{tabular}{||l|l|l||l|l||}
         \hline
         \hline
\multicolumn{1}{||c}{$\beta$}                               &
\multicolumn{1}{|c}{$c_{sw}$}                               &
\multicolumn{1}{|c||}{$\kappa$}                             &
\multicolumn{1}{c}{$M$}                                     &
\multicolumn{1}{|c||}{$\Delta M$}                           \\
         \hline
 8.00 & 1.40 & 0.12570 & \phantom{-}0.02742(25)    &\phantom{-}0.003117(207)  \\
 8.00 & 1.40 & 0.12630 & \phantom{-}0.007469(239)  &\phantom{-}0.002932(272)  \\
 8.00 & 1.40 & 0.12651 & \phantom{-}0.0001971(2329)& \phantom{-}0.002716(221) \\
 8.00 & 1.40 & 0.12680 & -0.009671(223)  & \phantom{-}0.003270(247)  \\
 8.00 & 1.40 & 0.12730 & -0.02596(28)    & \phantom{-}0.003221(256)  \\
         \hline
 8.00 & 1.60 & 0.12430 & \phantom{-}0.02266(23)    & -0.0004972(2019) \\
 8.00 & 1.60 & 0.12480 & \phantom{-}0.005679(245)  & -0.0008718(2676) \\
 8.00 & 1.60 & 0.12498 & \phantom{-}0.0002484(2335) & -0.0008608(2491) \\
 8.00 & 1.60 & 0.12520 & -0.007169(242)  & -0.0006004(2378) \\
 8.00 & 1.60 & 0.12570 & -0.02410(25)    & -0.001201(239)  \\
         \hline
 8.00 & 1.80 & 0.12240 & \phantom{-}0.03501(24)    & -0.003785(264)  \\
 8.00 & 1.80 & 0.12290 & \phantom{-}0.01858(26)    & -0.003763(179)  \\
 8.00 & 1.80 & 0.12344 & \phantom{-}0.0005959(2472) & -0.004154(247)  \\
 8.00 & 1.80 & 0.12350 & -0.002196(264)  & -0.005071(223)  \\
 8.00 & 1.80 & 0.12400 & -0.01861(27)    & -0.004060(270)  \\
         \hline
         \hline
 \end{tabular}
   \end{center}
\caption{$8^3\times 16$ results for $M$ and $\Delta M$ for $\beta = 8.00$.}
\label{table_run_b8p00}
\end{table}

\clearpage


\begin{table}[t]
   \begin{center}
      \begin{tabular}{||l|l|l||l|l||}
         \hline
         \hline
\multicolumn{1}{||c}{$\beta$}                               &
\multicolumn{1}{|c}{$c_{sw}$}                               &
\multicolumn{1}{|c||}{$\kappa$}                             &
\multicolumn{1}{c}{$M$}                                     &
\multicolumn{1}{|c||}{$\Delta M$}                           \\
         \hline
10.00 & 1.20 & 0.12570 & \phantom{-}0.01641(22)    &\phantom{-}0.003409(179)  \\
10.00 & 1.20 & 0.12619 & -0.0001306(1605) & \phantom{-}0.003338(182)  \\
10.00 & 1.20 & 0.12630 & -0.003541(173)  & \phantom{-}0.003321(217)  \\
10.00 & 1.20 & 0.12690 & -0.02350(20)    & \phantom{-}0.003296(198)  \\
10.00 & 1.20 & 0.12750 & -0.04340(17)    & \phantom{-}0.003247(206)  \\
         \hline
10.00 & 1.40 & 0.12410 & \phantom{-}0.03094(21)   &\phantom{-}0.0003442(1695) \\
10.00 & 1.40 & 0.12470 & \phantom{-}0.01351(29)   &\phantom{-}0.0006860(2239) \\
10.00 & 1.40 & 0.12507 & \phantom{-}0.0004563(4134)&\phantom{-}0.001032(171)  \\
10.00 & 1.40 & 0.12530 & -0.008549(319)  &\phantom{-}0.0005683(2022) \\
10.00 & 1.40 & 0.12590 & -0.02794(27)    &\phantom{-}0.001172(222)  \\
         \hline
10.00 & 1.60 & 0.12270 & \phantom{-}0.03342(46)    & -0.004086(267)  \\
10.00 & 1.60 & 0.12320 & \phantom{-}0.02152(16)    & -0.003744(145)  \\
10.00 & 1.60 & 0.12382 & \phantom{-}0.001171(165)  & -0.003759(157)  \\
10.00 & 1.60 & 0.12390 & -0.002455(294)  & -0.003601(186)  \\
10.00 & 1.60 & 0.12450 & -0.02161(19)    & -0.004090(163)  \\
         \hline
         \hline
 \end{tabular}
   \end{center}
\caption{$8^3\times 16$ results for $M$ and $\Delta M$ for $\beta = 10.00$.}
\label{table_run_b10p0}
\end{table}


\begin{table}[t]
   \begin{center}
      \begin{tabular}{||l|l|l||l|l||}
         \hline
         \hline
\multicolumn{1}{||c}{$\beta$}                               &
\multicolumn{1}{|c}{$c_{sw}$}                               &
\multicolumn{1}{|c||}{$\kappa$}                             &
\multicolumn{1}{c}{$M$}                                     &
\multicolumn{1}{|c||}{$\Delta M$}                           \\
         \hline
14.00 & 1.10 & 0.12500 & \phantom{-}0.01723(8)   & \phantom{-}0.002646(110) \\
14.00 & 1.10 & 0.12530 & \phantom{-}0.007452(89) & \phantom{-}0.002787(103) \\
14.00 & 1.10 & 0.12560 & -0.002273(87)           & \phantom{-}0.002941(114) \\
14.00 & 1.10 & 0.12590 & -0.01196(9)             & \phantom{-}0.002676(100) \\
14.00 & 1.10 & 0.12620 & -0.02194(9)             & \phantom{-}0.002684(113) \\
         \hline
14.00 & 1.20 & 0.12420 & \phantom{-}0.03218(36)  & \phantom{-}0.001696(167) \\
14.00 & 1.20 & 0.12470 & \phantom{-}0.01423(16)  & \phantom{-}0.001044(98)  \\
14.00 & 1.20 & 0.12530 & -0.002225(329)          & \phantom{-}0.002191(174) \\
14.00 & 1.20 & 0.12580 & -0.01786(46)            & \phantom{-}0.002320(199) \\
         \hline
14.00 & 1.30 & 0.12380 & \phantom{-}0.02900(34)  & -0.001514(170)   \\
14.00 & 1.30 & 0.12430 & \phantom{-}0.01132(40)  & -0.0004894(1641) \\
14.00 & 1.30 & 0.12480 & -0.005572(301)          & -0.0007392(1529) \\
14.00 & 1.30 & 0.12530 & -0.02027(107)           & -0.0009807(2838) \\
         \hline
         \hline
 \end{tabular}
   \end{center}
\caption{$8^3\times 16$ results for $M$ and $\Delta M$ for $\beta = 14.00$.}
\label{table_run_b14p0}
\end{table}

\clearpage


\begin{table}[t]
   \begin{center}
      \begin{tabular}{||l|l|l||l|l||}
         \hline
         \hline
\multicolumn{1}{||c}{$\beta$}                               &
\multicolumn{1}{|c}{$c_{sw}$}                               &
\multicolumn{1}{|c||}{$\kappa$}                             &
\multicolumn{1}{c}{$M$}                                     &
\multicolumn{1}{|c||}{$\Delta M$}                           \\
         \hline
 5.50 & 2.50 & 0.12300 &\phantom{-}0.02540(221) &\phantom{-}0.0006153(17988) \\
 5.50 & 2.50 & 0.12320 &\phantom{-}0.004367(3139)&\phantom{-}0.001051(1568) \\
 5.50 & 2.50 & 0.12335 & -0.002279(2162) &\phantom{-}0.001425(1272) \\
 5.50 & 2.50 & 0.12360 & -0.01981(151)   &\phantom{-}0.002050(1237) \\
         \hline
 5.50 & 2.60 & 0.12170 &\phantom{-}0.007744(2026) & -0.0009438(8558) \\
 5.50 & 2.60 & 0.12190 & -0.002810(1805) &\phantom{-}0.0002407(11134)\\
 5.50 & 2.60 & 0.12210 & -0.01117(179)   & -0.0008471(12790) \\
 5.50 & 2.60 & 0.12230 & -0.02560(168)   & -0.0007416(19293) \\
         \hline
 5.50 & 2.70 & 0.12015 &\phantom{-}0.01289(175)   & -0.0008967(10745) \\
 5.50 & 2.70 & 0.12040 &\phantom{-}0.001170(2838) & -0.0003062(16169) \\
 5.50 & 2.70 & 0.12070 & -0.01224(140)   & -0.001362(852)  \\
 5.50 & 2.70 & 0.12090 & -0.02138(153)   &\phantom{-}0.0002677(9645) \\
         \hline
 \end{tabular}
   \end{center}
\caption{$12^3\times 24$ results for $M$ and $\Delta M$ for $\beta = 5.50$.}
\label{table_run_5p50_12x24}
\end{table}


\begin{table}[t]
   \begin{center}
      \begin{tabular}{||l|l|l||l|l||}
         \hline
         \hline
\multicolumn{1}{||c}{$\beta$}                               &
\multicolumn{1}{|c}{$c_{sw}$}                               &
\multicolumn{1}{|c||}{$\kappa$}                             &
\multicolumn{1}{c}{$M$}                                     &
\multicolumn{1}{|c||}{$\Delta M$}                           \\
         \hline
 6.00 & 2.10 & 0.12430 &\phantom{-}0.01957(74) &\phantom{-}0.0003629(5316) \\
 6.00 & 2.10 & 0.12460 &\phantom{-}0.007496(680) &\phantom{-}0.0006202(5838) \\
 6.00 & 2.10 & 0.12495 & -0.001642(1038) &\phantom{-}0.001463(1070)  \\
 6.00 & 2.10 & 0.12520 & -0.01123(113)   &\phantom{-}0.0005411(5241) \\
         \hline
 6.00 & 2.20 & 0.12330 &\phantom{-}0.01228(67)    & -0.0008308(5383) \\
 6.00 & 2.20 & 0.12355 &\phantom{-}0.002046(917)  & -0.0008953(4855) \\
 6.00 & 2.20 & 0.12390 & -0.01153(83)    &\phantom{-}0.0005139(5375) \\
 6.00 & 2.20 & 0.12420 & -0.02019(76)    & -0.0003129(6525) \\
         \hline
 6.00 & 2.30 & 0.12190 &\phantom{-}0.02111(49)    & -0.001234(455)  \\
 6.00 & 2.30 & 0.12215 &\phantom{-}0.01067(68)    & -0.001233(833)  \\
 6.00 & 2.30 & 0.12240 &\phantom{-}0.002555(557)  & -0.0008735(5407) \\
 6.00 & 2.30 & 0.12280 & -0.01306(64)    & -0.0001565(5009) \\
         \hline
 6.00 & 2.40 & 0.12100 &\phantom{-}0.01273(49)    & -0.001217(461)  \\
 6.00 & 2.40 & 0.12120 &\phantom{-}0.005458(635)  & -0.002194(415)  \\
 6.00 & 2.40 & 0.12140 & -0.003718(533)  & -0.002257(514)  \\
 6.00 & 2.40 & 0.12160 & -0.009398(475)  & -0.001493(486)  \\
         \hline
 \end{tabular}
   \end{center}
\caption{$12^3\times 24$ results for $M$ and $\Delta M$ for $\beta = 6.00$.}
\label{table_run_6p00_12x24}
\end{table}

\clearpage



\end{document}